\documentclass[lettersize,journal]{IEEEtran}
\usepackage{amsmath,amssymb,amsfonts}
\usepackage{algorithmic}
\usepackage{algorithm}
\usepackage{array}
\usepackage[caption=false,font=normalsize,labelfont=sf,textfont=sf]{subfig}
\usepackage{textcomp}
\usepackage{stfloats}
\usepackage{url}
\usepackage{verbatim}
\usepackage{graphicx}
\usepackage{cite}
\usepackage{textcomp}
\usepackage{xcolor}
\usepackage{listings}

\newtheorem{thm}{\protect\theoremname}
\newtheorem{lem}[thm]{\protect\lemmaname}
\newtheorem{defn}[thm]{Definition}
\providecommand{\lemmaname}{Lemma}

\lstdefinelanguage{Julia}%
  {morekeywords={abstract,begin,break,case,catch,const,continue,do,else,elseif,end,export,false,for,function,immutable,import,importall,if,in,macro,module,otherwise,quote,return,switch,true,try,type,typealias,using,while},%
   sensitive=true,%
   alsoother={\$},%
   morecomment=[l]\#,%
   morecomment=[n]{\#=}{=\#},%
   morestring=[s]{"}{"},%
   morestring=[m]{'}{'},%
}[keywords,comments,strings]%

\begin{document}

\title{On The Performance of Prefix-Sum Parallel Kalman Filters and Smoothers on GPUs}

\author{Simo S\"arkk\"a,~\IEEEmembership{Senior Member,~IEEE,}
\and
\'Angel F. Garc\'ia-Fern\'andez
\thanks{Simo Särkkä (simo.sarkka@aalto.fi) is with Dept.
        of Electrical Engineering and Automation, Aalto University, Finland, and \'Angel F. Garc\'ia-Fern\'andez (angel.garcia.fernandez@upm.es) is with IPTC, ETSI Telecomunicaci\'on, Universidad Polit\'ecnica de Madrid, Spain.}%
}

\markboth{Journal of \LaTeX\ Class Files,~Vol.~14, No.~8, August~2021}%
{Shell \MakeLowercase{\textit{et al.}}: A Sample Article Using IEEEtran.cls for IEEE Journals}

\IEEEpubid{0000--0000/00\$00.00~\copyright~2021 IEEE}

\maketitle

\begin{abstract}
This paper presents an experimental evaluation of parallel-in-time Kalman filters and smoothers using graphics processing units (GPUs). In particular, the paper evaluates different all-prefix-sum algorithms, that is, parallel scan algorithms for temporal parallelization of Kalman filters and smoothers in two ways: by calculating the required number of operations via simulation, and by measuring the actual run time of the algorithms on real GPU hardware. In addition, a novel parallel-in-time two-filter smoother is proposed and experimentally evaluated. Julia code for Metal and CUDA implementations of all the algorithms is made publicly available. 
\end{abstract}


\section{Introduction}
\IEEEPARstart{K}{alman} filters and smoothers \cite{Kalman:1960, Rauch65, Fraser:1969, Jazwinski:1970} are classical algorithms for state estimation in stochastic dynamic systems. They are widely used in target tracking, process control, mobile computing, and other sensor fusion applications \cite{Bar-Shalom+Li+Kirubarajan:2001, Sarkka+Svensson:2023}. The classical Kalman filter and smoother algorithms \cite{Kalman:1960, Rauch65, Fraser:1969, Jazwinski:1970, Bar-Shalom+Li+Kirubarajan:2001, Sarkka+Svensson:2023} are sequential algorithms that loop over the measurement data (or length $T$) in forward and backward directions, respectively. Although they have linear time complexity $O(T)$ and are time-optimal in the sequential sense, they are suboptimal on parallel hardware such as graphics processing units (GPUs). Furthermore, in many applications, such as sensor fusion systems in mobile phones, it would be desirable to run the Kalman filters and smoothers on the GPU, thereby freeing the main central processing unit (CPU) for other tasks. Unfortunately, Kalman filters and smoothers in their classical form are very slow and inefficient in terms of resource utilization when run on a GPU.

S\"arkk\"a \& Garc\'ia-Fern\'andez \cite{Sarkka:2021} developed parallel-in-time algorithms for Kalman filtering and smoothing, which provide a solution to the problem above. The algorithms convert the sequential computation of $O(T)$ steps into $O(\log T)$ parallel steps, which utilize the parallel GPU hardware more effectively. The algorithms are based on reformulating the filtering and smoothing problems as the computation of generalized prefix sums for suitably defined associative operators. The ideal time complexity $O(\log T)$ results from the span complexity of the underlying prefix-sum (i.e., scan) algorithm \cite{Blelloch:1989, Blelloch:1990}, which, though, is only realized in practice when the number of computational cores is high enough. These algorithms enable the execution of Kalman filters and smoothers on GPUs significantly faster, utilizing the GPU's parallel computational resources more effectively. Alternatively, or additionally, it would be possible to parallelize the Kalman filter equations themselves \cite{Lee+Haykin:1993, Lyster:1997, Rosen:2013}, but in this paper, the focus is on the temporal parallelization.

Although the reference \cite{Sarkka:2021} only analyzed the computational advantage of the parallel algorithms by using simulated parallel hardware, supplemental JAX \cite{Bradbury:2018} and TensorFlow \cite{Abadi_et_al:2015} implementations for GPUs were later provided\footnote{\url{https://github.com/EEA-sensors/sequential-parallelization-examples/tree/main/python/temporal-parallelization-bayes-smoothers}}. Furthermore, the algorithms, or actually their extensions, have been experimentally evaluated on GPUs in subsequent references \cite{Yaghoobi:2021, Yaghoobi:2025} using the JAX framework \cite{Bradbury:2018}.

Although the aforementioned experimental results already demonstrate the advantage of temporal parallelization of Kalman filters and smoothers, the experimental results are still limited. The experiments were implemented using the relatively high-level JAX \cite{Bradbury:2018} and TensorFlow  \cite{Abadi_et_al:2015} frameworks, and the prefix-sum algorithms used were the default implementations provided by the frameworks. The results still leave it uncertain what the overhead of the frameworks themselves is and what the effect of the underlying prefix-sum algorithm is on the practical GPU performance. Also, the impact of the particular GPU choice remains unknown. In this paper, the aim is to fill these gaps by evaluating different prefix-sum algorithms on two different GPUs using a lower-level cross-platform implementation in Julia \cite{Bezanson:2017}.

Another limitation in the reference \cite{Sarkka:2021} and the subsequent works is that they are based on the so-called Rauch--Tung--Striebel (RTS) formulation \cite{Rauch65, Sarkka+Svensson:2023} of the smoother. The RTS formulation has the disadvantage that the backward pass depends on the forward pass, and hence, they cannot be performed simultaneously, even when multiple GPUs are available. An alternative formulation of the smoother is the so-called two-filter smoother \cite{Fraser:1969}, which has independent forward and backward passes that can be run simultaneously on two GPUs. Although such two-filter smoothing in the context of hidden Markov models (HMMs) was discussed in \cite{Hassan:2021} and an analogous parallel linear quadratic control algorithm was developed in \cite{Sarkka:2023}, this kind of algorithm has not yet appeared in the literature. In this paper, we develop the parallel two-filter smoother and test its performance in a two-GPU setup.

As discussed above, the selection of the underlying prefix-sum algorithm also affects the performance of the parallel Kalman filter and smoother algorithms. Although, for example, JAX and TensorFlow frameworks \cite{Bradbury:2018, Abadi_et_al:2015} use the associative scan algorithm of Blelloch \cite{Blelloch:1990} for parallel prefix sum computation, there exist various alternative prefix sum methods. An early example is the Hillis--Steele algorithm \cite{Hillis:1986}, even earlier is the Ladner \& Fischer circuit \cite{Ladner:1980, Karp:1990}, and various alternative and improved algorithms have been proposed \cite{Sengupta:2006, Sengupta:2007, Dotsenko:2008, Merrill:2009}. Reviews of prefix-sum algorithms, historical remarks, and discussion on their performance in CPU and GPU contexts can be found in \cite{Merrill:2009, Breshears:2009}. The prefix-sum methods have various other applications, including sorting, recurrence equations, computer graphics, and optimal control \cite{Hillis:1986, Blelloch:1989, Blelloch:1990, Blelloch:1990book, Kohlhoff:2013, Sarkka:2023}.

The contributions of the paper are:
\begin{itemize}
\item Analysis of the number of floating point operations required by Kalman filtering/smoothing methods with different prefix-sum algorithms.
\item Experimental evaluation of Kalman filters and smoothers on two GPUs using dedicated Julia Metal.jl/CUDA.jl implementations with different all-prefix-sum algorithms.
\item Two-filter smoother version of the parallel Kalman smoother and its evaluation using 2 GPUs.
\item We also provide an open source code for the implementation of the algorithms.
\end{itemize}

The structure of the paper is the following. Section \ref{sec:parallel_review} reviews relevant parallel all-prefix sum algorithms. Section \ref{sec:sequential_Kalman} reviews sequential Bayesian and Kalman filtering and smoothing algorithms. Section \ref{sec:parallel_Kalman} reviews the parallel versions of these algorithms and presents a novel two-filter form of the Bayesian and Kalman smoother. The implementations of the algorithms in Julia are explained in Section \ref{sec:implementation}. Experimental results are analyzed in Section \ref{sec:experiments}. Finally, conclusions are drawn in Section \ref{sec:conclusion}.

\section{Parallel all-prefix-sum algorithms} \label{sec:parallel_review}

This section reviews relevant parallel all-prefix sum algorithms required for the parallel implementation of filters and smoothers. The input size of the algorithms is denoted as $T$, which is also the number of measurements in the corresponding state estimation problem. All the algorithms presented assume that $T$ is a power of $2$, but they can be easily generalized to an arbitrary $T$ by padding the series of elements with neutral elements.

\subsection{Complexity measures of parallel programs}

In the following sections, we use simplified forms of parallel random access machine (PRAM) models \cite{JaJa:1992}, where the memory access is assumed to take negligible (i.e., zero) computational time and a single application of the associative operator $\otimes$ in a single processor takes a single computational time unit. We use two different complexity measures. The first one is the {\em span complexity}, $\mathrm{span}(T)$, which refers to the parallel computing steps taken by the program on a parallel computer with an unlimited number of processors. The second one is {\em work complexity}, $\mathrm{work}(T)$, which is the total amount of computations taken by the algorithm. We assume the latter to be independent of the number of processors. We also always have $\mathrm{span}(T) \le \mathrm{work}(T)$.

The actual computational time, $\mathrm{time}(T, P)$, taken by the algorithm also depends on the number of processors $P$ in the computer. We always have $\mathrm{time}(T, P) \ge \mathrm{span}(T)$ with the equality attained when $P \to \infty$. Therefore, with a large number of processors relative to $T$, the span complexity determines the computational time. However, we also have the work complexity bound $\mathrm{time}(T,P) \ge \mathrm{work}(T) / P$, and thus, with a small number of processors relative to $T$, we always end up following the work complexity. With growing $T$, the faster $\mathrm{work}(T)$ grows with $T$, the faster the latter complexity regime is reached. A useful way to analyze the algorithms is to take $P = T$ and hence analyze how $\mathrm{work}(T) / T$ grows. If it is in $O(1)$, then the algorithm scales quite well; otherwise, we can expect to run out of processors quite quickly.

\subsection{Parallel computation of all-prefix-sums}

All-prefix-sums for general operators can be used as computational primitives in various algorithms, including sorting, recurrence equations, and computer graphics, as discussed in \cite{Hillis:1986, Blelloch:1989}. Additionally, they are applicable in state estimation and optimal control, as explored in \cite{Sarkka:2021, Sarkka:2023, Hassan:2021, Yaghoobi:2021, Yaghoobi:2025}. In the all-prefix-sums problem, we are given a series of $T$ elements $a_1, a_2, \ldots, a_T$ and a binary associative operator $\otimes$ defined on the elements. The all-prefix-sum or scan operation computes $s_1=a_1$, $s_2=a_1 \otimes a_2$, \ldots, $s_T = a_1 \otimes a_2 \otimes \cdots \otimes a_T$. We can compute the all-prefix-sum operation sequentially in $O(T)$ time by using a simple loop. However, more importantly for this paper, it is also possible to compute it in parallel in $O(\log(T))$ span time with algorithms explained in the rest of this section. It is worth noting that the span complexity of $O(\log(T))$ only applies in the limit of an infinite number of processing cores, and the actual complexity on a fixed number of cores also heavily depends on the work complexity of the method at hand.

We also often need to compute the reversed all-prefix-sums $\bar{s}_T=a_T$, $\bar{s}_{T-1} = a_{T-1} \otimes a_T$, \ldots, $\bar{s}_1 = a_1 \otimes a_2 \otimes \cdots \otimes a_T$. Given a forward scan algorithm, we can compute these, for example, by reversing the series before and after applying the scan algorithm (reversion is a span $O(1)$ parallel operation). An alternative, often a more efficient implementation, involves reversing the indices inside the algorithm itself. The reversed prefix sums are needed in the computation of the Kalman smoothing solutions.

In the two-filter smoother that we introduce in Section \ref{subsec:tf_smoother}, we combine the results of forward and reversed all-prefix sums. The idea is that, as we have $a_1 \otimes a_2 \otimes \cdots \otimes a_T = (a_1 \otimes a_2 \otimes \cdots \otimes a_{k-1}) \otimes (a_k \otimes a_{k+1} \otimes \cdots \otimes a_T)$, then we also have $s_T = s_{k-1} \otimes \bar{s}_k$ for all $k$, which is useful in the two-filter smoother.

\subsection{Hillis--Steele algorithm}

The Hillis--Steele algorithm \cite{Hillis:1986} is a simple parallel prefix sum algorithm that has $O(\log T)$ span complexity. The algorithm is based on recursively summing the first half of the series into the second half, as shown in Algorithm~\ref{alg:hillis-steele}. Its disadvantage is the high work complexity of $O(T \log T)$, which thus causes the time complexity with a finite number of processors to be super-linear. Hence $\mathrm{work}(T)/ T \in O(\log T)$, which hints that the algorithm runs out of processors quite quickly. We indeed see this problem later on in the experiments.

\begin{algorithm}
        \textbf{Input:} The elements $\{ a_k \}_{k=1}^T$ and associative operator $\otimes$.\\
        \textbf{Output:} The result in $\{ a_k \}_{k=1}^T$.
        \begin{algorithmic}[1]
            \FOR{$d\gets0$ \textbf{to} $\log_2 T$}
                \STATE $b \gets \mathrm{copy}(a)$ // Take a copy of the current series
                \FOR[Compute in parallel]{$i\gets1$ \textbf{to} $T-2^d$}
                    \STATE $j \gets i + 2^d$
                    \STATE $a_j \gets b_i \otimes b_j$
                \ENDFOR
            \ENDFOR
        \end{algorithmic}
        \caption{Parallel-scan algorithm of Hillis and Steele \cite{Hillis:1986}.}
        \label{alg:hillis-steele}
\end{algorithm}

\begin{figure}[htbp]
\centerline{\includegraphics[width=\columnwidth]{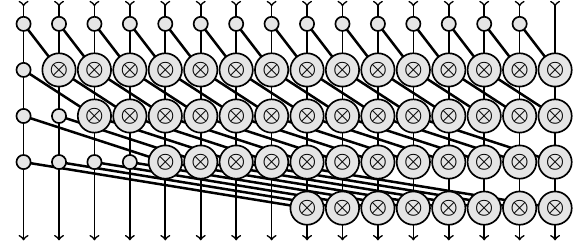}}
\caption{Illustration of the operation of the algorithm of Hillis and Steele (Alg.~\ref{alg:hillis-steele}) \cite{Hillis:1986} with 16 elements. Each column is a data storage element with $\otimes$ being the associative operator between two elements. The arrows indicate increasing time. At the final time step, the data elements contain the prefix sums.}
\label{fig:circuit-hillissteele}
\end{figure}

The operation of the Hillis--Steele \cite{Hillis:1986} algorithm is illustrated in Fig.~\ref{fig:circuit-hillissteele}. The algorithm is in-place in the sense that, after computing the current row from the previous row, data from the earlier rows is no longer needed. However, in practice, we need to use double buffering (or a copy of the previous row, as in Alg.~\ref{alg:hillis-steele}), because otherwise we would overwrite the data before using it in the computation. Thus, in that sense, the required storage is twice the length of the input array, $2T$.

\subsection{Blelloch's algorithm}

The algorithm of Blelloch \cite{Blelloch:1990} is based on an up-sweep and a down-sweep, which can be seen as in-place implementations of upward and downward traversals in a binary tree. The algorithm is presented as Algorithm~\ref{alg:blelloch}. The span complexity of the method is also $O(\log T)$, but its work complexity is $O(T)$, which is better than that of the Hillis--Steele algorithm. We indeed have $\mathrm{work}(T) / T \in O(1)$, which suggests that the method runs out of processors more slowly than Hillis--Steele, and this phenomenon is also observed in the experiments. Thus, the algorithm is much faster in practice than Hillis--Steele. It is worth noting that the method requires the storage of length $2T$ -- this is because we need additional storage for the copy of the original array (for the final pass). In fact, the temporary variables $t$ need storage space as well, so in that sense the total storage space is actually $2T + T/2$.

\begin{algorithm}
        \textbf{Input:} The elements $\{ a_k \}_{k=1}^T$ and associative operator $\otimes$.\\
        \textbf{Output:} The result in $\{ a_k \}_{k=1}^T$.
        \begin{algorithmic}[1]
                \STATE $b \gets \mathrm{copy}(a)$ // Store the input
                \STATE // Up-sweep:
                \FOR{$d\gets0$ \textbf{to} $\log_2 T - 1$}
                \FOR[Compute in parallel]{$i\gets0$ \textbf{to} $T-1$ \textbf{by} $2^{d+1}$}
                \STATE $j \gets i + 2^d$
                \STATE $k \gets i + 2^{d+1}$
                \STATE $a_k \gets a_j \otimes a_k$
                \ENDFOR
                \ENDFOR
                \STATE $a_T \leftarrow 0$ \COMMENT{Here, $0$ is the neutral element for $\otimes$}
                \STATE // Down-sweep:
                \FOR{$d\gets\log_2 T - 1$ \textbf{to} $0$}
                \FOR[Compute in parallel]{$i\gets0$ \textbf{to} $T-1$ \textbf{by} $2^{d+1}$}
                \STATE $j \gets i + 2^d$
                \STATE $k \gets i + 2^{d+1}$
                \STATE $t \gets a_j$
                \STATE $a_j \gets a_k$
                \STATE $a_k \gets a_k \otimes t$
                 \ENDFOR
                \ENDFOR
                \STATE // Final pass to form the inclusive scan:
                \FOR[Compute in parallel]{$i\gets1$ \textbf{to} $T$}
                \STATE $a_i \gets a_i \otimes b_i$
                \ENDFOR
        \end{algorithmic}
        \caption{Parallel-scan algorithm of Blelloch \cite{Blelloch:1990}.}
        \label{alg:blelloch}
\end{algorithm}

\begin{figure}[htbp]
\centerline{\includegraphics[width=\columnwidth]{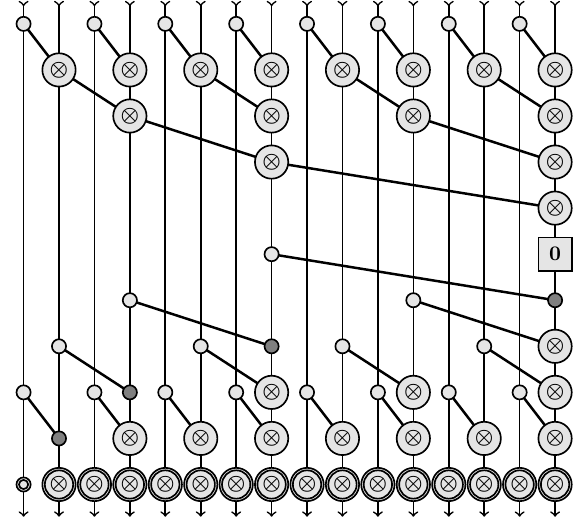}}
\caption{Illustration of the operation of Blelloch's algorithm (Alg.~\ref{alg:blelloch}) \cite{Blelloch:1990} with 16 elements.}
\label{fig:circuit-blelloch}
\end{figure}

The operation of Blelloch's algorithm \cite{Blelloch:1990} is illustrated in Fig.~\ref{fig:circuit-blelloch}. In addition to the need for a temporary variable, a special feature of the algorithm is the step that involves only zeroing the last element in the array. Furthermore, because the algorithm initially computes the exclusive scan, the last step (final pass) consists of a fully parallel operation between the original array and the exclusive scan result.

\subsection{Ladner's and Fischer's circuit}

Already in 1980, Ladner and Fischer \cite{Ladner:1980} published an approach to form logarithmic-depth circuits for computing prefix sums. In algorithmic terms, the basic idea \cite{Karp:1990} is first to compute the pairs $\hat{a}_i = a_{2i-1} \otimes a_{2i}$ and to compute the prefix sums of the resulting array recursively. An in-place implementation of the method is presented as Algorithm~\ref{alg:inplace-lafi}. Although this type of implementation is known to parallel computing experts, this particular algorithm is difficult to find in existing literature. Its advantage is that it is genuinely in-place, as no additional storage is required beyond the original array of length $T$. The algorithm has a span complexity of $O(\log T)$ and work complexity $O(T)$. As Blelloch's algorithm, this method also has $\mathrm{work}(T) / T \in O(1)$; however, it always requires two parallel steps fewer than Blelloch's algorithm, which is also a practical advantage.

\begin{algorithm}
        \textbf{Input:} The elements $\{ a_k \}_{k=1}^T$ and associative operator $\otimes$.\\
        \textbf{Output:} The result in $\{ a_k \}_{k=1}^T$.
        \begin{algorithmic}[1]
                \STATE // Up-sweep:
                \FOR{$d\gets0$ \textbf{to} $\log_2 T - 1$}
                \FOR[Compute in parallel]{$i\gets0$ \textbf{to} $T-1$ \textbf{by} $2^{d+1}$}
                \STATE $j \gets i + 2^d$
                \STATE $k \gets i + 2^{d+1}$
                \STATE $a_k \gets a_j \otimes a_k$
                \ENDFOR
                \ENDFOR
                \STATE // Down-sweep:
                \FOR{$d\gets\log_2 T - 1$ \textbf{to} $0$}
                \FOR[In parallel]{$i\gets2^{d+1}$ \textbf{to} $T-1$ \textbf{by} $2^{d+1}$}
                \STATE $j \gets i + 2^d$
                \STATE $a_j \gets a_i \otimes a_j$
                \ENDFOR
                \ENDFOR
        \end{algorithmic}
        \caption{In-place implementation of the parallel scan circuit of Ladner and Fischer \cite{Ladner:1980}.}
        \label{alg:inplace-lafi}
\end{algorithm}

\begin{figure}[htbp]
\centerline{\includegraphics[width=\columnwidth]{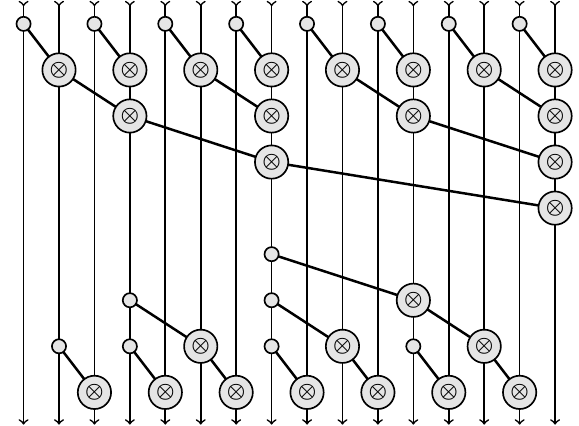}}
\caption{Illustration of the operation of the in-place Ladner and Fischer's algorithm (Alg.~\ref{alg:inplace-lafi}) \cite{Ladner:1980} with 16 elements.}
\label{fig:circuit-inplace-lafi}
\end{figure}

The operation of Algorithm~\ref{alg:inplace-lafi} is illustrated in Fig.~\ref{fig:circuit-inplace-lafi}. Unlike Blelloch's algorithm, there is no zeroing of elements, and the method computes the inclusive scan directly.

\subsection{Hybrid algorithms of Sengupta et al.}

The algorithm of Sengupta et al.~\cite{Sengupta:2006} can be seen to employ a tree traversal similar to the Blelloch method, along with an additional fallback to Hillis--Steele for processing short series (say, $\le N$) within the algorithm execution. The time complexity of the method is $O(\log T)$, and the work complexity is $O(T)$. Because the Hillis--Steele algorithm has a smaller absolute number of span steps, this can, in principle, make the algorithm faster. However, as will be seen in the experiments, this does not always happen in practice, due to higher work complexity. Therefore, a practical special case of the algorithm is also obtained by setting $N = 1$, in which case it becomes essentially equivalent to Algorithm~\ref{alg:inplace-lafi} but with a different storage arrangement for the intermediate values.

\begin{algorithm}
        \textbf{Input:} The elements $\{ a_k \}_{k=1}^T$, associative operator $\otimes$, threshold parameter $N$.\\
        \textbf{Output:} The result in $\{ a_k \}_{k=1}^T$.
        \begin{algorithmic}[1]
            \STATE $a^{(0)} \gets \mathrm{copy}(a)$ // The zeroth level is the plain series
            \STATE $d^* \gets \log_2 (T / N)$ // Up to which level we process
            \FOR{$d\gets1$ \textbf{to} $d^*$}
                \STATE // Reduce pass:
                \FOR[Compute in parallel]{$i\gets1$ \textbf{to} $T/2^{d}$}
                    \STATE $a_i^{(d)} \gets a_{2i-1}^{(d-1)} \otimes a_{2i}^{(d-1)}$
                \ENDFOR
	        \ENDFOR
            \STATE \COMMENT{Run Hillis--Steele algorithm on array $a^{(d^*)}$.} 
            \STATE // Down-sweep:
            \FOR{$d\gets d^*-1$ \textbf{to} $0$}
                \FOR[Compute in parallel]{$i\gets1$ \textbf{to} $T/2^{d}$}
                    \IF{$i > 1$}
                        \IF{$i~\textrm{mod}~2 = 1$}
                            \STATE $a_i^{(d)} \gets a_{(i-1)/2}^{(d+1)} \otimes a_i^{(d)}$
                        \ELSE
                            \STATE $a_i^{(d)} \gets a_{i/2}^{(d+1)}$
                        \ENDIF
                    \ELSE
                        \STATE $a_i^{(d)} \gets a_i^{(d)}$
                    \ENDIF
                \ENDFOR
            \ENDFOR
        \STATE $a \gets a^{(0)}$ // The final result
        \end{algorithmic}
        \caption{Parallel-scan algorithm of Sengupta et al.~\cite{Sengupta:2006}.}
        \label{alg:sengupta}
\end{algorithm}

\subsection{Other algorithms}

There also exist various modifications to the methods described above, based on block processing and other ideas \cite{Sengupta:2007, Dotsenko:2008, Merrill:2009}. Most of these modifications, however, can be applied equally to all the algorithms above and therefore should not significantly alter their relative performance.

\section{Sequential Bayesian and Kalman filtering and smoothing} \label{sec:sequential_Kalman}

In this paper, the focus is on parallel implementations of Kalman filters and smoothers, which are special cases of Bayesian filters and smoothers (see, e.g., \cite{Sarkka+Svensson:2023}). We now review these in a sequential setting. Bayesian filters and smoothers are sequential algorithms for estimating the state of a process $x_k \in \mathbb{R}^{n_{x}}$ at time step $k$ from noisy measurements $y_k \in \mathbb{R}^{n_{y}}$. In particular, they aim at computing the conditional (posterior) distributions of the states given the measurements. The state evolves according to a Markov process with transition density $p\left(x_{k+1} \mid x_{k}\right)$. At each time step, the process is observed via a noisy measurement with density $p\left(y_{k} \mid x_{k}\right)$.

\subsection{Bayesian filtering} \label{sec:bayesian-filtering}

In Bayesian filtering, we aim compute the posterior density $p(x_k \mid y_{1:k})$ of the state $x_k$ given the measurements $ y_{1:k}=(y_1,\ldots,y_k)$ up to time step $k$. The filtering problem can be solved with linear complexity $O(T)$ using the Bayesian filtering recursion, which consists of the prediction and update steps given by \cite{Ho+Lee:1964, Sarkka+Svensson:2023}
\begin{equation}
    p(x_k \mid y_{1:k-1})
    = \int p(x_k \mid x_{k-1})
       \, p(x_{k-1} \mid y_{1:k-1}) \, \mathrm{d}x_{k-1},
\label{eq:bayesf_p}
\end{equation}
\begin{equation}
    p(x_k \mid y_{1:k})
    = \frac{p(y_k \mid x_k)
       \, p(x_k \mid y_{1:k-1})}
       {\int p(y_k \mid x_k)
          \, p(x_k \mid y_{1:k-1}) \, \mathrm{d}x_k}.
\label{eq:bayesf_u}
\end{equation}
In practice, we start from a prior distribution $p(x_0) \triangleq p(x_0 \mid y_{1:0})$ and perform the above operations for $k = 1,2,3,\ldots,T$, where $T$ is the total number of measurements.

Kalman filter \cite{Kalman:1960} is a special case of the above Bayesian filter, for linear Gaussian systems.

\subsubsection{Kalman filtering}
We consider the linear-Gaussian models of the form
\begin{align}
p\left(x_{k}\mid x_{k-1}\right) & =\mathrm{N}\left(x_{k};F_{k-1}x_{k-1}+u_{k-1},Q_{k-1}\right),\label{eq:transition_density_Kalman}\\
p\left(y_{k}\mid x_{k}\right) & =\mathrm{N}\left(y_{k};H_{k}x_{k}+d_{k},R_{k}\right),\label{eq:measurement_density_Kalman}
\end{align}
where $F_{k-1}\in\mathbb{R}^{n_{x}\times n_{x}}$ and $H_{k}\in\mathbb{R}^{n_{y}\times n_{x}}$
are known matrices, $u_{k-1}\in\mathbb{R}^{n_{x}}$ and $d_{k}\in\mathbb{R}^{n_{y}}$
are known vectors, and $Q_{k-1}\in\mathbb{R}^{n_{x}\times n_{x}}$
and $R_{k}\in\mathbb{R}^{n_{y}\times n_{y}}$ are covariance matrices. The prior at time step 0 is $p\left(x_{0}\right) = \mathrm{N}\left(x_{0};\overline{x}_{0|0},P_{0|0}\right)$.

In this case, the filtering density is Gaussian
\begin{align}
p(x_{k}\mid y_{1:k}) & = \mathrm{N}\left(x_{k};\overline{x}_{k|k},P_{k|k}\right).
\end{align}
The filtering mean $\overline{x}_{k|k}$ and covariance matrix $P_{k|k}$ are calculated via the Kalman filtering recursion with $O(T)$ complexity. The prediction step computes
\begin{align}
\overline{x}_{k|k-1} & =F_{k-1}\overline{x}_{k-1|k-1}+u_{k-1},\label{eq:KF_mean_pred}\\
P_{k|k-1} &= F_{k-1}P_{k-1|k-1}F_{k-1}^{\top}+Q_{k-1}.
\end{align}
The update step computes
\begin{align}
S_{k} & =H_{k}P_{k|k-1}H_{k}^{\top}+R_{k}, \\
\overline{x}_{k|k} & =\overline{x}_{k|k-1}+P_{k|k-1}H_{k}^{\top}S_{k}^{-1}\left(y_{k}-H_{k}\overline{x}_{k|k-1}-d_{k}\right),\\
P_{k|k} & =P_{k|k-1}-P_{k|k-1}H_{k}^{\top}S_{k}^{-1}H_{k}P_{k|k-1}.\label{eq:KF_cov_upd}\
\end{align}

The Kalman filter algorithm amounts to starting from the prior mean and covariance $\overline{x}_{0|0},P_{0|0}$ and performing the above prediction and update steps for $k = 1,2,3,\ldots,T$.

\subsection{Smoothing} \label{sec:bayesian-smoothing}

In smoothing, we compute the density $p(x_k \mid y_{1:T})$ for $k<T$, given the measurements up to a time step $T$. The smoothing problem can also be solved with linear complexity $O(T)$ by running an additional backward recursion, either in a forward-backward form, or via two-filter smoothing.

\subsubsection{Forward-backward smoother}
Given the filtering densities for $k=1,\ldots,T$, the forward-backward smoother uses the following recursion for $k=T-1,\ldots,1$ \cite{Kitagawa:1987, Sarkka+Svensson:2023}:
\begin{equation}
\begin{split}
  &p(x_k \mid y_{1:T}) \\
  &= p(x_k \mid y_{1:k})
     \int \frac{p(x_{k+1} \mid x_k)
     	\, p(x_{k+1} \mid y_{1:T})}
     {p(x_{k+1} \mid y_{1:k})}
         \mathrm{d} x_{k+1}.
\end{split}
\label{eq:bsmoothingeqs}
\end{equation}

In the linear/Gaussian model, described by \eqref{eq:transition_density_Kalman} and \eqref{eq:measurement_density_Kalman}, the forward pass corresponds to the Kalman filter. The backward pass computes the smoothed mean $\overline{x}_{k|T}$ and covariance matrix $P_{k|T}$ via the following recursion from $k=T-1$ to $k=1$:
\begin{align}
G_{k} &= P_{k|k}F_{k}^{\top}\left(P_{k+1|k}\right)^{-1}, \\ 
\overline{x}_{k|T} &=\overline{x}_{k|k}+G_{k}\left(\overline{x}_{k+1|T}-\overline{x}_{k+1|k}\right),\\
P_{k|T} &= P_{k|k}+G_{k}\left(P_{k+1|T}-P_{k+1|k}\right)G_{k}^{\top}.
\end{align}
This smoothing algorithm is referred to as the Rauch--Tung--Striebel RTS smoother \cite{Rauch65}.

\subsubsection{Two-filter smoother}
The two-filter smoother is based on the factorization \cite{Kitagawa94}
\begin{align}
p(x_{k}\mid y_{1:T}) & \propto p(x_{k} \mid y_{1:k}) \, p(y_{k+1:T}\mid x_{k}). \label{eq:two_filter_factor}
\end{align}

The two-filter smoother consists of two independent recursions, one running forward, which corresponds to the Bayesian filtering recursion in \eqref{eq:bayesf_p} and \eqref{eq:bayesf_u} to calculate the first factor, and one running backwards to calculate the second factor, given by
\begin{align}
p(y_{k+1:T}\mid x_{k}) & =\int p\left(x_{k+1}\mid x_{k}\right)p(y_{k+1:T}\mid x_{k+1})dx_{k+1},\\
p(y_{k:T}\mid x_{k}) & = p(y_{k+1:T}\mid x_{k}) \, p\left(y_{k}\mid x_{k}\right).
\end{align}

For the linear Gaussian case, whose model is given by \eqref{eq:transition_density_Kalman} and \eqref{eq:measurement_density_Kalman}, the second factor in \eqref{eq:two_filter_factor} is of the form
\begin{align}
p(y_{k+1:T}\mid x_{k}) & \propto\mathrm{N}_{I}\left(x_{k};\eta_{k|k+1:T},J_{k|k+1:T}\right)
\end{align}
which represents a Gaussian density in information form with information vector $\eta_{k|k+1:T}$ and information matrix $J_{k|k+1:T}$, evaluated at $x_k$.

The backward filter recursion starts with \cite{Fraser:1969}
\begin{align}
\eta_{T|T:T} & =H_{T}^{\top}R_{T}^{-1}\left(y_{T}-d_{T}\right), \\
J_{T|T:T} & =H_{T}^{\top}R_{T}^{-1}H_{T}.
\end{align}
The backward prediction step computes \cite{Kitagawa_arxiv23}

\begin{align}
\eta_{k-1|k:T}&=F_{k-1}^{\top}\left(I+J_{k|k:T}Q_{k-1}\right)^{-1}\left(\eta_{k|k:T}-J_{k|k:T}u_{k-1}\right) \label{eq:tf_vector_pred}\\
J_{k-1|k:T}&=F_{k-1}^{\top}\left(I+J_{k|k:T}Q_{k-1}\right)^{-1}J_{k|k:T}F_{k-1}.
\end{align}
The backward prediction step can also be seen as the result of combining two filtering elements, which will be introduced in Section \ref{subsec:parallel_Kalman}, when there is no measurement at time step $k-1$.

The update step computes
\begin{align}
\eta_{k-1|k-1:T} & =\eta_{k-1|k:T}+H_{k-1}^{\top}R_{k-1}^{-1}\left(y_{k-1}-d_{k-1}\right),\\
J_{k-1|k-1:T} & =J_{k-1|k:T}+H_{k-1}^{\top}R_{k-1}^{-1}H_{k-1}.\label{eq:tf_matrix_upd}
\end{align}
 
Using \eqref{eq:two_filter_factor}, the smoothed density is then Gaussian with mean and covariance matrix 
\begin{align}
\overline{x}_{k|T} & =\left(I+P_{k|k}J_{k|k+1:T}\right)^{-1}\left(\overline{x}_{k|k}+P_{k|k}\eta_{k|k+1:T}\right), \label{eq:mean_tf_combination}\\
P_{k|T} & =\left(I+P_{k|k}J_{k|k+1:T}\right)^{-1}P_{k|k}.\label{eq:cov_tf_combination}
\end{align}

In summary, for the linear Gaussian case, the two-filter smoother runs the Kalman filter forward, using \eqref{eq:KF_mean_pred}-\eqref{eq:KF_cov_upd}, and, possibly in parallel, it runs the backward filter, using \eqref{eq:tf_vector_pred}-\eqref{eq:tf_matrix_upd}. The smoothed means and covariance matrices for all time steps are then obtained with \eqref{eq:mean_tf_combination} and \eqref{eq:cov_tf_combination}.

\section{Parallel Bayesian and Kalman filtering and smoothing} \label{sec:parallel_Kalman}

In the previous section, we briefly reviewed sequential algorithms for Bayesian filtering and smoothing, as well as those for Kalman filtering and smoothing. In this section, the aim is to review parallel Bayesian/Kalman filtering methods and the RTS-type of smoothers first proposed in \cite{Sarkka:2021}. Finally, in Sections~\ref{subsec:tf_smoother} and \ref{subsec:tf_Gaussian_smoother}, we describe a novel two-filter form of Bayesian/Kalman smoother.

\subsection{Parallel Bayesian filtering} \label{sub:Parallel_filt}

The parallel formulation of Bayesian smoothing in \cite{Sarkka:2021} is based on first computing the Bayesian filtering solution in parallel and then the Bayesian smoothing solution. In this section, we review the parallelization of the Bayesian filter. The parallelization is based on the use of parallel prefix-sum algorithms (see Sec.~\ref {sec:parallel_review}), which require that the corresponding operator is associative. However, the operator corresponding to a single time step in the Bayesian filtering equations \eqref{eq:bayesf_p} and \eqref{eq:bayesf_u} is not associative. It can be made associative as follows.

An element used for filtering $a_{k}\in\mathcal{F}$ is of the form
$a_{k}=\left(f,g\right)$ where $f$ is
a two-variable function $f:\mathbb{R}^{n_{x}}\times\mathbb{R}^{n_{x}}\rightarrow\mathbb{R}$
and $g$ is a one-variable function $g:\mathbb{R}^{n_{x}}\rightarrow\mathbb{R}$. 

\begin{defn}
\label{def:Operator_filtering}Given two elements $\left(f_{i},g_{i}\right)\in\mathcal{F}$
and $\left(f_{j},g_{j}\right)\in\mathcal{F}$, the associative operator
$\otimes$ for Bayesian filtering is \cite{Sarkka:2021}
\begin{equation}
\left(f_{i},g_{i}\right)\otimes\left(f_{j},g_{j}\right)  =\left(f_{i,j},g_{i,j}\right),
\end{equation}
where
\begin{equation}
\begin{split}
f_{i,j}\left(x,z\right) & =\frac{\int g_{j}\left(y\right)f_{j}\left(x,y\right)f_{i}\left(y,z\right)dy}{\int g_{j}\left(y\right)f_{i}\left(y,z\right)dy}, \\
g_{i,j}\left(z\right) & =g_{i}\left(z\right)\int g_{j}\left(y\right)f_{i}\left(y,z\right)dy.
\end{split}
\end{equation}
\end{defn}

Specifically, for $k>1$ the element $a_k$ is \cite{Sarkka:2021}
\begin{align} \label{eq:element_filtering}
a_{k} & = \begin{pmatrix}
p(x_{k} \mid y_{k},x_{k-1})\\
p(y_{k} \mid  x_{k-1})
\end{pmatrix},
\end{align}
where $f_k(x_k,x_{k-1})=p(x_{k} \mid y_{k},x_{k-1})$ and $g_k(x_{k-1})=p(y_{k} \mid  x_{k-1})$. Furthermore, we have 
\begin{align} \label{eq:element_filtering_1}
a_{1}	&=\begin{pmatrix}
p(x_{1}\mid y_{1})\\
p(y_{1})
\end{pmatrix}.
\end{align}

These definitions of element and associative operator yield
\begin{align} \label{eq:filtering_prefix_sums}
	a_{1}\otimes a_{2}\otimes\cdots\otimes a_{k} &= \begin{pmatrix}
	p\left(x_{k}\mid y_{1:k}\right)\\
	p\left(y_{1:k}\right)
	\end{pmatrix}.
\end{align}

Let us understand this combination rule in more detail. Let $i$,
$j$, and $k$ be three time steps such that $i<j<k$, and $a_{i,j}=a_{i}\otimes a_{i+1}\otimes\ldots\otimes a_{j}$.
Then, we have that
\begin{align}
a_{i,j} &= \begin{pmatrix}
p(x_{j}\mid y_{i+1:j},x_{i})\\
p(y_{i+1:j}\mid x_{i})
\end{pmatrix}, \\
a_{j,k} &= \begin{pmatrix}
p(x_{k}\mid y_{j+1:k},x_{j})\\
p(y_{j+1:k}\mid x_{j})
\end{pmatrix}.
\end{align}

The combination rule therefore computes $a_{i,k}=a_{i,j}\otimes a_{j,k}$ using
the following relations, which can be obtained from the Markov properties
of the model,
\begin{align}
p(x_{k}\mid y_{i+1:j},x_{j}) & \propto\int p(y_{j+1:k}\mid x_{j})p(x_{k}\mid y_{j+1:k},x_{j}), \nonumber \\
 & \quad\times p(x_{j}\mid y_{i+1:j},x_{i})dx_{j}\\
p(y_{i+1:k}\mid x_{i}) & =p(y_{i+1:j}\mid x_{i})\nonumber \\
 & \quad\times\int p(y_{j+1:k}\mid x_{j})p(x_{j}\mid y_{i+1:j},x_{i})dx_{j}.
\end{align}
Computing $a_1 \otimes a_2 \otimes \cdots \otimes a_k$ sequentially is equivalent to computing the filtering solution using equations \eqref{eq:bayesf_p} and \eqref{eq:bayesf_u}, and additionally, the normalization constant $p(y_{1:k})$ using a separate recursion. However, the operator is associative, and hence we can parallelize it using parallel prefix sums.

\subsection{Parallel Bayesian smoothing}

The parallel smoother derived in \cite{Sarkka:2021} has the so-called Rauch--Tung--Striebel form corresponding to parallelization of the equation \eqref{eq:bsmoothingeqs}. In this formulation, the element used for smoothing $a_{k}\in\mathcal{S}$ is a two-variable
function of the form $a:\mathbb{R}^{n_{x}}\times\mathbb{R}^{n_{x}}\rightarrow\mathbb{R}$.
\begin{defn}
\label{def:Operator_smoothing}Given two elements $a_{i}\in\mathcal{S}$
and $a_{j}\in\mathcal{S}$, the associative operator $\otimes$ for
Bayesian smoothing is
\begin{align*}
a_{i}\otimes a_{j} & =a_{i,j},
\end{align*}
where
\begin{align*}
a_{i,j}\left(x,z\right) & =\int a_{i}\left(x,y\right)a_{j}\left(y,z\right)dy.
\end{align*}
\end{defn}

In particular, the element that enables parallel computation of Bayesian smoothing
is \cite{Sarkka:2021}
\begin{align}
a_{k} & =p(x_{k}\mid y_{1:k},x_{k+1})
\end{align}
with $a_{T}=p(x_{T}\mid y_{1:T})$.

These definitions of element and associative operator lead to
\begin{align}
a_{k}\otimes a_{k+1}\otimes\cdots\otimes a_{T} & =p(x_{k}\mid y_{1:T}).
\end{align}
As in filtering, let us understand this combination rule in more detail.
Let $i$, $j$, and $k$ be three time steps such that $i<j<k$, the
elements are then
\begin{align}
a_{i,j} &= p(x_{i}\mid y_{1:j-1},x_{j}), \\
a_{j,k} &=p(x_{j}\mid y_{1:k-1},x_{k}).
\end{align}
The associative operator combines these results by computing
\begin{align}
a_{i,k} & =\int p(x_{i}\mid y_{1:j-1},x_{j})p(x_{j}\mid y_{1:k-1},x_{k})dx_{j}\\
 & =p(x_{i}\mid y_{1:k-1},x_{k}).
\end{align}
This result is direct from the Markov properties of the model.

\subsection{Parallel Kalman filter} \label{subsec:parallel_Kalman}

Let us consider the linear-Gaussian system, with model in \eqref{eq:transition_density_Kalman} and \eqref{eq:measurement_density_Kalman}. To solve the associated filtering problem with $O (\log T)$ complexity, the element $a_k$ in \eqref{eq:element_filtering} is described by
\begin{equation}
\begin{split}
	p\left(x_{k}\mid y_{k},x_{k-1}\right)&=\mathrm{N}\left(x_{k};A_{k}x_{k-1}+b_{k},C_{k}\right), \\
	p\left(y_{k}\mid x_{k-1}\right)&\propto\mathrm{N}_{I}\left(x_{k-1};\eta{}_{k},J_{k}\right). \\
\end{split}
\end{equation}
The parameters $\left(A_{k},b_{k},C_{k},\eta_{k},J_{k}\right)$ are given as follows \cite{Sarkka:2021}. For $k > 1$, $\left(A_{k},b_{k},C_{k} \right)$ are
\begin{equation}
\begin{split}
	S_{k} & =H_{k}Q_{k-1}H_{k}^{\top}+R_{k}, \\
    K_{k} & =Q_{k-1}H_{k}^{\top}S_{k}^{-1}, \\
    A_{k} & =\left(I_{n_{x}}-K_{k}H_{k}\right)F_{k-1}, \\
	b_{k} & =u_{k-1}+K_{k}\left(y_{k}-H_{k}u_{k-1}-d_{k}\right), \\
	C_{k} & =\left(I_{n_{x}}-K_{k}H_{k}\right)Q_{k-1}. \\
\end{split}
\label{eq:pkf_init1}
\end{equation}
For $k = 1$, $\left(A_{1},b_{1},C_{1} \right)$ are
\begin{equation}
\begin{split}
   \overline{x}_{1|0} &= F_{0} \overline{x}_{0|0} + u_{0}, \\
   P_{1|0} &= F_{0} P_{0|0} F^\top_{0} + Q_{0}, \\
   S_1 &= H_1 P_{1|0} H_1^\top + R_1, \\
   K_1 &= P_{1|0} H^\top_1 S_1^{-1}, \\
   A_1 &= 0, \\
   b_1 &= \overline{x}_{1|0} + K_1 [y_1 - H_1 \overline{x}_{1|0} - d_1], \\
   C_1 &= P_{1|0} - K_1 S_1 K_1^\top.
\end{split}
\label{eq:pkf_init2}
\end{equation}
	The parameters $(\eta_{k}, J_k)$ of the second term are given as
\begin{equation}
\begin{split}
	\eta_{k} & =F_{k-1}^{\top}H_{k}^{\top}S_{k}^{-1}\left(y_{k}-H_{k}u_{k-1}-d_{k}\right), \\
	J_{k} & =F_{k-1}^{\top}H_{k}^{\top}S_{k}^{-1}H_{k}F_{k-1},
\end{split}
\label{eq:pkf_init3}
\end{equation}
for $k=1,\ldots,T$.

How the associative operator for filtering works when the elements are defined in this manner is provided in the following lemma.
\begin{lem}
	\label{prop:Operator_linear_filtering}Given two elements $\left(A_{i},b_{i},C_{i},\eta_{i},J_{i}\right)$ and $\left(A_{j},b_{j},C_{j},\eta_{j},J_{j}\right)$ the binary operator $\otimes$ for filtering returns an element $\left(A_{i,j},b_{i,j},C_{i,j},\eta_{i,j},J_{i,j}\right)$ where
	\begin{equation}
    \begin{split}
	A_{i,j} & =A_{j}\left(I_{n_{x}}+C_{i}J_{j}\right)^{-1}A_{i},\\
	b_{i,j} & =A_{j}\left(I_{n_{x}}+C_{i}J_{j}\right)^{-1}\left(b_{i}+C_{i}\eta_{j}\right)+b_{j},\\
	C_{i,j} & =A_{j}\left(I_{n_{x}}+C_{i}J_{j}\right)^{-1}C_{i}A_{j}^{\top}+C_{j},\\
	\eta_{i,j} & =A_{i}^{\top}\left(I_{n_{x}}+J_{j}C_{i}\right)^{-1}\left(\eta_{j}-J_{j}b_{i}\right)+\eta_{i},\\
	J_{i,j} & =A_{i}^{\top}\left(I_{n_{x}}+J_{j}C_{i}\right)^{-1}J_{j}A_{i}+J_{i}.
	\end{split}
\end{equation}
\end{lem}

The pseudocode of the parallel Kalman filter is provided in Algorithm \ref{alg:pkf}.

\begin{algorithm}
	\textbf{Input:} The measurements $y_{1:T}$, the model matrices $F_{0:T-1}$, $Q_{0:T-1}$, $H_{1:T}$, $R_{1:T}$, inputs $u_{0:T-1}$, and initial statistics $\overline{x}_{0|0}, P_{0|0}$. \\
	\textbf{Output:} The Kalman filter means and covariances $\overline{x}_{k\mid k}$ and $P_{k \mid k}$ for $k = 1,\ldots,T$. 
    \begin{algorithmic}[1]
		\STATE // Initialization: 
    	\FOR[Compute in parallel]{$k\gets 1$ \textbf{to} $T$}
    		\STATE Compute initial element $$a_k \gets (A_k, b_k, C_k, \eta_k, J_k)$$ using \eqref{eq:pkf_init1}, \eqref{eq:pkf_init2}, and \eqref{eq:pkf_init3}.
		\ENDFOR
		\STATE // Associative scan: 
		\STATE Define filtering operator $\otimes$ via Lemma~\ref{prop:Operator_linear_filtering}.
		\STATE Call parallel associative scan to obtain the prefix sums: $$s_{1:T} = \textrm{parallel\_scan}(a_{1:T}, \otimes).$$
		\STATE // Extract the results: 
    	\FOR[Compute in parallel]{$i\gets 1$ \textbf{to} $T$}
			\STATE Unpack the prefix sums: $(A_k^+, b_k^+, C_k^+, \eta_k^+, J_k^+) = s_k$.
			\STATE Extract the Kalman filter mean and covariance:
			\begin{equation}
			\begin{split}
				\overline{x}_{k \mid k} &\gets b_k^+, \\
				P_{k \mid k} &\gets C_k^+.
			\end{split}
			\end{equation}
		\ENDFOR
	\end{algorithmic}
	\caption{Parallel Kalman filter from \cite{Sarkka:2021}.}
	\label{alg:pkf}
\end{algorithm}

\subsection{Parallel Rauch--Tung--Striebel smoother}

 Let us now formulate the Rauch--Tung--Striebel (RTS) smoother \cite{Rauch65} in a parallel form. For the linear-Gaussian model in  \eqref{eq:transition_density_Kalman} and \eqref{eq:measurement_density_Kalman}, the element for forward-backward smoothing becomes 
\begin{equation}  
   p(x_{k}\mid y_{1:k},x_{k+1}) = \mathcal{N}\left(x_{k};E_{k}x_{k+1}+g_{k},L_{k}\right),
   \label{eqn:conditional_probability}
\end{equation}
where the parameters $\left(E_{k},g_{k},L_{k}\right)$ are given as follows. For $k < T$
\begin{equation}
    \begin{split}
	E_{k} & =P_{k|k}F_{k}^{\top}\left(F_{k}P_{k|k}F_{k}^{\top}+Q_{k}\right)^{-1},\\
	g_{k} & =\overline{x}_{k|k}-E_{k}\left(F_{k}\overline{x}_{k|k}+u_{k}\right),\\
	L_{k} & =P_{k|k}-E_{k}F_{k}P_{k|k},
\end{split}
\label{eq:prts_init1}
\end{equation}
and, for $k=T$,
\begin{equation}
    \begin{split}
	E_{T} & = 0, \\
	g_{T} & = \overline{x}_{T|T},\\
	L_{T} & = P_{T|T}.
\end{split}
\label{eq:prts_init2}
\end{equation}

The following lemma describes how the associative operator for smoothing works when the elements are defined as above.

\begin{lem}
	\label{prop:Operator_linear_smoothing}Given two elements $\left(E_{i},g_{i},L_{i}\right)$ and $\left(E_{j},g_{j},L_{j}\right)$ the associative operator $\otimes$ for smoothing returns an element $\left(E_{i,j},g_{i,j},L_{i,j}\right)$ where
	\begin{equation}
    \begin{split}
	   E_{i,j} & =E_{i}E_{j},\\
	   g_{i,j} & =E_{i}g_{j}+g_{i},\\
	   L_{i,j} & =E_{i}L_{j}E_{i}^{\top}+L_{i}.
	\end{split}
\end{equation}
\end{lem}

The pseudocode of the parallel RTS smoother is provided in Algorithm \ref{alg:prts}.

\begin{algorithm}
	\textbf{Input:} The Kalman filter results $\overline{x}_{k \mid k}$ and $P_{k \mid k}$ for $k = 1,\ldots,T$, the model matrices $F_{0:T-1}$, $Q_{0:T-1}$, and inputs $u_{0:T-1}$. \\
	\textbf{Output:} The RTS smoother means and covariances $\overline{x}_{k\mid T}$ and $P_{k \mid T}$ for $k = 1,\ldots,T$. 
    \begin{algorithmic}[1]
		\STATE // Initialization: 
    	\FOR[Compute in parallel]{$k\gets 1$ \textbf{to} $T$}
    		\STATE Compute initial element $$a_k \gets (E_k, g_k, L_k)$$ using \eqref{eq:prts_init1} and \eqref{eq:prts_init2}.
		\ENDFOR
		\STATE // Associative scan: 
		\STATE Define smoothing operator $\otimes$ via Lemma~\ref{prop:Operator_linear_smoothing}.
		\STATE Call backward parallel associative scan to obtain the prefix sums: $$\overline{s}_{1:T} = \textrm{reverse\_parallel\_scan}(a_{1:T}, \otimes).$$
		\STATE // Extract the results: 
    	\FOR[Compute in parallel]{$i\gets 1$ \textbf{to} $T$}
			\STATE Unpack the prefix sums: $(E_k^+, g_k^+, L_k^+) = \overline{s}_k$.
			\STATE Extract the RTS smoother mean and covariance:
			\begin{equation}
			\begin{split}
				\overline{x}_{k \mid T} &\gets g_k^+, \\
				P_{k \mid T} &\gets L_k^+.
			\end{split}
			\end{equation}
		\ENDFOR
	\end{algorithmic}
	\caption{Parallel RTS smoother from \cite{Sarkka:2021}.}
	\label{alg:prts}
\end{algorithm}

\subsection{Parallel two-filter smoother} \label{subsec:tf_smoother}

The novel two-filter smoother that we propose in this paper is based on the factorization of the smoothed density in \eqref{eq:two_filter_factor}. The first factor in \eqref{eq:two_filter_factor}, $p(x_{k} \mid y_{1:k})$, can be computed for all $k$ using parallel computation using \eqref{eq:filtering_prefix_sums}. In addition, using the filtering element in \eqref{eq:element_filtering} and its associative operator, it is direct to check that
\begin{align}
a_{k+1}\otimes \cdots \otimes a_{T} & = \begin{pmatrix}
p(x_{T} \mid y_{k+1:T},x_{k})\\
p(y_{k+1:T} \mid x_{k})
\end{pmatrix}.
\end{align}
Therefore, the second factor in \eqref{eq:two_filter_factor}, $p(y_{k+1:T} \mid x_{k})$,  can be computed for all $k$ using the same elements and associative operator as for filtering, but using a reversed all-prefix-sums operation.  In practice, to align the elements of the forward and backward filters better, it is sometimes beneficial to use the equivalent formulation
\begin{align}
a_{k+1}\otimes \cdots \otimes a_{T} \otimes e & = \begin{pmatrix}
p(x_{T} \mid y_{k+1:T},x_{k})\\
p(y_{k+1:T} \mid x_{k})
\end{pmatrix},
\end{align}
where $e$ is the neural element for $\otimes$. 

Once we have run the forward and backward filters in parallel, which can be executed on two different GPUs, we can combine their results to obtain the smoothed density using \eqref{eq:two_filter_factor}. This combination can be done in parallel for all time steps. When operations are performed on two different GPUs, the initialization must also be performed twice to avoid sending the initialization results from one GPU to another. This is also what we do in our implementation.

\subsection{Parallel two-filter linear-Gaussian smoother} \label{subsec:tf_Gaussian_smoother}

This section explains the implementation of the parallel two-filter smoother in Section \ref{subsec:tf_smoother} for the linear-Gaussian models in \eqref{eq:transition_density_Kalman} and \eqref{eq:measurement_density_Kalman}. In this case,  we run the parallel Kalman filter using the elements and associative operator in Section \ref{subsec:parallel_Kalman}, collecting the filtering mean $\overline{x}_{k|k}$ and covariance matrix $P_{k|k}$ for all $k$. 

In parallel to this operation (for instance, using a second GPU),  we run a reverse all-prefix-sums operation, using the same elements and associative operator, collecting the backward information vector $\eta_{k|k+1:T}$ and information matrix $J_{k|k+1:T}$ for all $k$. Finally, we can obtain the smoothed mean $\overline{x}_{k|T}$ and covariance matrix $P_{k|T}$ for all $k$ in parallel using \eqref{eq:mean_tf_combination} and \eqref{eq:cov_tf_combination}. The pseudocode of the parallel two-filter smoother is provided in Algorithm \ref{alg:ptf}.

\begin{algorithm}
	\textbf{Input:} The measurements $y_{1:T}$, the model matrices $F_{0:T-1}$, $Q_{0:T-1}$, $H_{1:T}$, $R_{1:T}$, inputs $u_{0:T-1}$, and initial statistics $\overline{x}_{0|0}, P_{0|0}$. \\
	\textbf{Output:} The smoother means and covariances $\overline{x}_{k\mid T}$ and $P_{k \mid T}$ for $k = 1,\ldots,T$. 
    \begin{algorithmic}[1]
		\STATE // Initialization: 
    	\FOR[Compute in parallel]{$k\gets 1$ \textbf{to} $T$}
    		\STATE Compute initial element $$a_k \gets (A_k, b_k, C_k, \eta_k, J_k)$$ using \eqref{eq:pkf_init1}, \eqref{eq:pkf_init2}, and \eqref{eq:pkf_init3}.
		\ENDFOR
		\STATE // Associative scan: 
		\STATE Define filtering operator $\otimes$ via Lemma~\ref{prop:Operator_linear_filtering}.
		\STATE Call parallel associative scan to obtain the prefix sums: $$s_{1:T} = \textrm{parallel\_scan}(a_{1:T}, \otimes).$$
		\STATE Call reverse parallel associative scan to obtain the reversed prefix sums (this can be done in parallel with the above): $$\overline{s}_{1:T} = \textrm{reverse\_parallel\_scan}((a_{2:T},e), \otimes),$$
        where $e = (0,0,0,0,0)$.
		\STATE // Extract the results: 
    	\FOR[Compute in parallel]{$i\gets 1$ \textbf{to} $T$}
			\STATE Unpack the prefix sums: 
			\begin{equation}
			\begin{split}
				(A_k^+, b_k^+, C_k^+, \eta_k^+, J_k^+) \gets s_k. \\
				(A_k^-, b_k^-, C_k^-, \eta_k^-, J_k^-) \gets \overline{s}_k. \\
			\end{split}
			\end{equation}
			
			\STATE Extract the forward and backward Kalman filter results and compute the smoothing solution:
			\begin{equation}
			\begin{split}
				\overline{x}_{k \mid k} &\gets b_k^+, \\
				P_{k \mid k} &\gets C_k^+, \\
				\eta_{k \mid k+1:T} &\gets \eta_k^-, \\
				J_{k \mid k+1:T} &\gets J_k^-, \\
				\overline{x}_{k|T} &\gets \left(I+P_{k|k}J_{k|k+1:T}\right)^{-1}\left(\overline{x}_{k|k}+P_{k|k}\eta_{k|k+1:T}\right), \\
				P_{k|T} &\gets \left(I+P_{k|k}J_{k|k+1:T}\right)^{-1}P_{k|k}.
				\end{split}
			\end{equation}
		\ENDFOR
	\end{algorithmic}
	\caption{Parallel two-filter smoother, including the Kalman filter pass.}
	\label{alg:ptf}
\end{algorithm}

\section{Implementation of the Algorithms} \label{sec:implementation}

In this section, the aim is to discuss how the algorithms described in the previous sections have been implemented for the experimental evaluation. All the implementations are available in \verb+(link to be revealed upon acceptance)+.\footnote{A zip file containing the implementations is provided as a supplementary file for the peer review.}

\subsection{Julia's Metal.jl and CUDA.jl}

Our implementations have been done purely in the Julia programming language \cite{Bezanson:2017} and, in particular, using Julia's GPU compiler interface packages \cite{besard2018juliagpu, besard2019prototyping} for Apple Metal (Metal.jl) \cite{JuliaMetal:2025} and NVIDIA's CUDA (CUDA.jl) \cite{JuliaCUDA:2025}. We implemented the core matrix routines required by the Kalman filters and smoothers in platform-independent Julia, which is compiled for the GPUs using these packages or for CPU using the standard (just-in-time) compilation features of Julia. 

The program listings in Figs.~\ref{fig:uppass_kernel} and \ref{fig:index_stride_f} illustrate the principle of how the code is implemented. The code in Fig.~\ref{fig:uppass_kernel} is an illustrative platform-independent implementation of the $d$th level of the up-sweep, which appears in Algorithms~\ref{alg:blelloch} and \ref{alg:inplace-lafi} -- the actual Julia code is slightly more complicated to avoid copying of values around, but the principle is the same. The platform-dependent part is given as a closure \texttt{index\_stride\_f} which returns the index and stride for the stride loop. The associative operator $\otimes$ is provided as a closure \texttt{op}. The elements themselves are given in array \texttt{elems}, the level in $d$, and the total number of elements in $T$. 

\begin{figure}[htbp]
\begin{lstlisting}
function upsweep_kernel(index_stride_f, op, elems, d, T)
    index, stride = index_stride_f()

    delta1 = 1 << d
    delta2 = 1 << (d+1)
    range_j = ((index - 1) * delta2 + delta1):
              (stride * delta2):(T - 1 + delta1)
    range_k = ((index - 1) * delta2 + delta2):
              (stride * delta2):(T - 1 + delta2)

    for (j,k) in zip(range_j, range_k)
        elems[k] = op(elems[j], elems[k])
    end
    return
end
\end{lstlisting}
\caption{Sketch of Julia implementation of generic up-sweep GPU kernel of a parallel associative scan.}
\label{fig:uppass_kernel}
\end{figure}

Fig.~\ref{fig:index_stride_f} shows example implementations of the index and stride computation for Metal and CUDA. A similar implementation can also be done for the CPU. It is worth noting that the Julia GPU Compiler \cite{besard2018juliagpu, besard2019prototyping} itself supports similar platform-abstractions over GPUs, but we have chosen to use this one to allow for more flexible benchmarking with CPU.

\begin{figure}[htbp]
\begin{lstlisting}
function index_stride_f_metal()
    index = thread_position_in_grid_1d()
    stride = threads_per_grid_1d()
end
\end{lstlisting}
\begin{lstlisting}
function index_stride_f_cuda()
    index = (blockIdx().x-1) * blockDim().x + threadIdx().x
    stride = gridDim().x * blockDim().x
end
\end{lstlisting}
\caption{Functions to compute the index and stride on Metal and CUDA.}
\label{fig:index_stride_f}
\end{figure}

\subsection{Implementation of the parallel prefix-sum algorithms}

All the parallel prefix-sum (i.e., scan) algorithms have been implemented as platform-independent codes resembling the one shown in Fig.~\ref{fig:uppass_kernel}. One practical difference, though, is that to avoid dynamic memory allocation, the operator directly stores the result in the array instead of using the assignment operator $=$ for it. Hence, the associative operation is implemented in the form $\texttt{fun!(i, j, k, elems)}$, which performs the operation $\texttt{elems[i] = op(elems[j], elems[k])}$. More precisely, $\texttt{elems}$ is actually a tuple of arrays in order to support the associative operations required in the Kalman filters and smoothers.

\subsection{Implementation of parallel Kalman filters and smoothers}

For the Kalman filters and smoothers, we implemented a custom set of primitive matrix operations (addition, multiplication, Cholesky, LU, triangular solving, etc.) that can be run in the GPU kernels. The parallel Kalman filter and smoothers were implemented using these primitives with the help of the associative scan algorithms described in the previous section. 

\subsection{Calculating the theoretical number of operations} \label{sec:flop-counting}

To count the floating point operations required by the algorithms, we implemented flop-counting on top of the matrix routines using Julia's operator overloading, and made a simulated index and stride interface for it (as in Fig.~\ref{fig:index_stride_f}), which counts the flops for a given number of simulated threads. This allowed us to count the number of operations using exactly the same platform-independent code as was run on the GPUs. 

\section{Experimental results} \label{sec:experiments}

In this section, we report the experimental results consisting of counting the number of floating-point operations for simulated runs of the methods, and the algorithm speed measurements using Apple Metal GPU and NVIDIA CUDA GPU, along with the CPUs of the computers. 

\subsection{Experimental setup}

The performance was measured by using a model of the form given in Eqs.~\eqref{eq:transition_density_Kalman} and \eqref{eq:measurement_density_Kalman} with 4-dimensional state and 2-dimensional measurements such that $F_k$, $u_k$, $Q_k$, $H_k$, $d_k$, and $R_k$ were randomly generated for each $k$ separately. They were generated from multivariate unit Gaussian distributions, but the $F_k$ matrix was replaced with $0.99$ times its Q factor from QR-factorization to ensure numerical stability, and $Q_k$ and $R_k$ were formed as products of the random matrices $\Xi \, \Xi^\top$, where $\Xi$ is a Gaussian matrix, to ensure that they are positive definite. The initial mean $m_1$ and covariance $P_1$ were similarly randomly generated. The measurements were generated by simulating data from the model.

In the GPU and CPU speed experiments, the time was measured with Julia's \verb+@elapsed+ macro, and the time was computed from 12 runs, where the first two were discarded (to eliminate the effect of JIT compilation), and the time was estimated to be the median of the rest of the runs. All the computations were performed using 32-bit floating-point numbers.

The experiments were run on two computers:
\begin{itemize}
\item {\bf Metal computer:} Apple MacBook Pro laptop with Apple M3 Max GPU (30 Cores) and with Apple M3 Max GPU at 2400 MHz.
\item {\bf CUDA computer:} Desktop computer with 4 NVIDIA A100-SXM4-80GB GPUs and AMD EPYC 7643 48-Core Processor at 1500 MHz.
\end{itemize}

The associative scan methods were the following:
\begin{itemize}
\item {\bf Hillis--Steele:} This is the Hillis--Steele parallel associative scan in Algorithm~\ref{alg:hillis-steele}.
\item {\bf Blelloch:} This is the Blelloch's parallel associative scan in Algorithm~\ref{alg:blelloch}.
\item {\bf Inplace LaFi:} This is the inplace implementation of the Ladner--Fischer circuit given in Algorithm~\ref{alg:inplace-lafi}.
\item {\bf Sengupta A:} This the Sengupta's parallel associative scan in Algorithm~\ref{alg:sengupta} with $N=1$.
\item {\bf Sengupta B:} This the Sengupta's parallel associative scan in Algorithm~\ref{alg:sengupta} with $N$ set of $14750$ in the Metal experiments, $20000$ in CUDA experiments, and $15000$ in simulated hardware experiments. The Metal and CUDA $N$'s were empirically selected to give good performance.
\end{itemize}

The parallel Kalman filtering and smoothing methods are the following:
\begin{itemize}
\item {\bf PKF:} This is the parallel Kalman filter described in Algorithm \ref{alg:pkf}.
\item {\bf PRTS:} This is the parallel Rauch--Tung--Striebel smoother (including the parallel Kalman filter pass) described in Algorithm \ref{alg:prts}.
\item {\bf PTFS:} This is the parallel two-filter smoother (including the parallel Kalman filter pass) described in Algorithm \ref{alg:ptf}.
\end{itemize}

\subsection{Performance on a simulated parallel hardware} \label{sec:simulated}

As described in Section~\ref{sec:flop-counting}, we implemented floating point operation counting on top of the methods, which can be used to run the codes with a simulated number of threads and estimate the time taken by the methods (``time'' in terms of number of floating point operations). In the experiments, we simulated a system with a GPU with 15000 threads, which is slightly more than the number of threads supported by the Metal computer's GPU and somewhat less than the CUDA computer's GPU.

\begin{figure}[htbp]
\centerline{\includegraphics[width=\columnwidth]{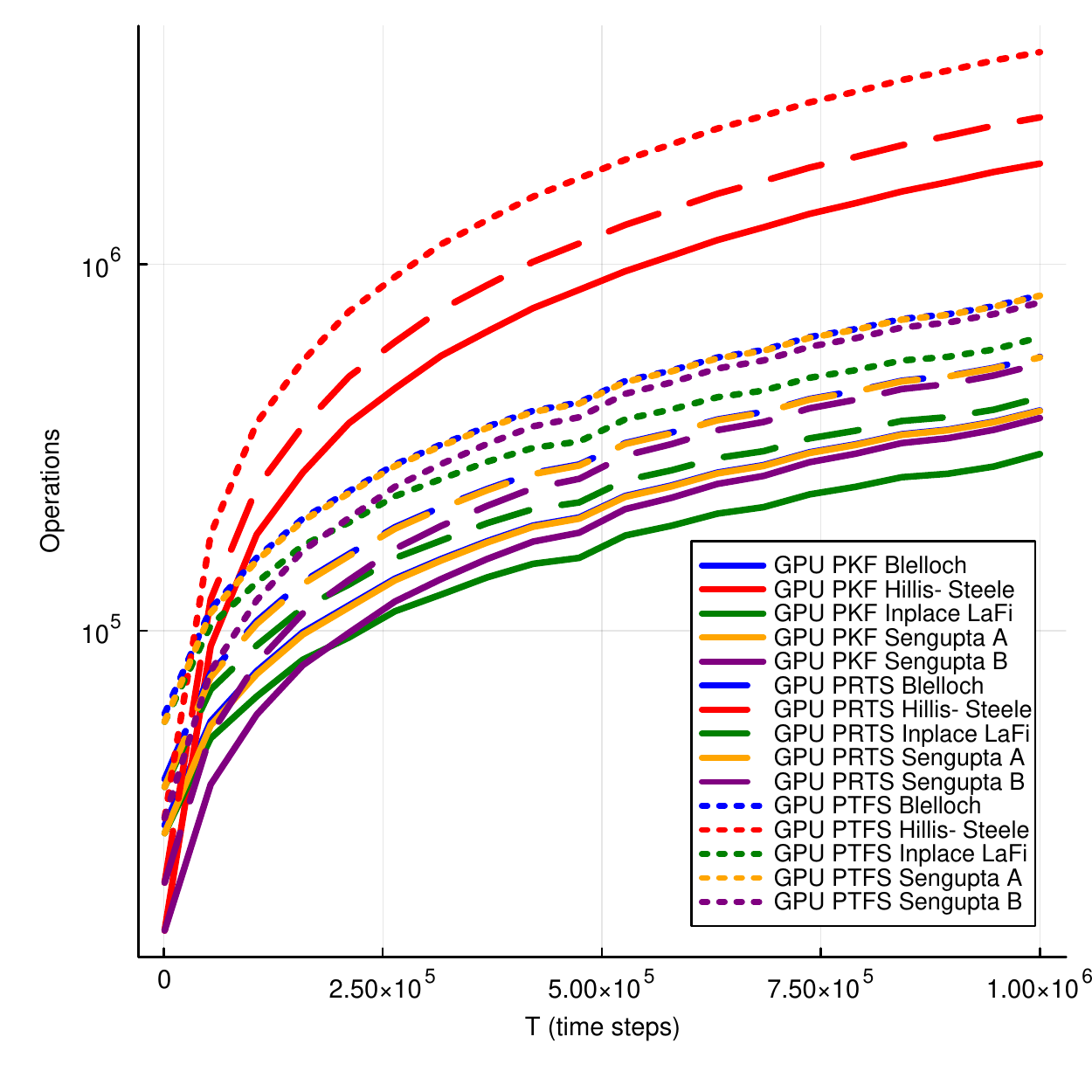}}
\caption{Number of operations of the different algorithm combinations as a function of time series length $T$.}
\label{fig:pkfs_results_flops}
\end{figure}

\begin{figure}[htbp]
\centerline{\includegraphics[width=\columnwidth]{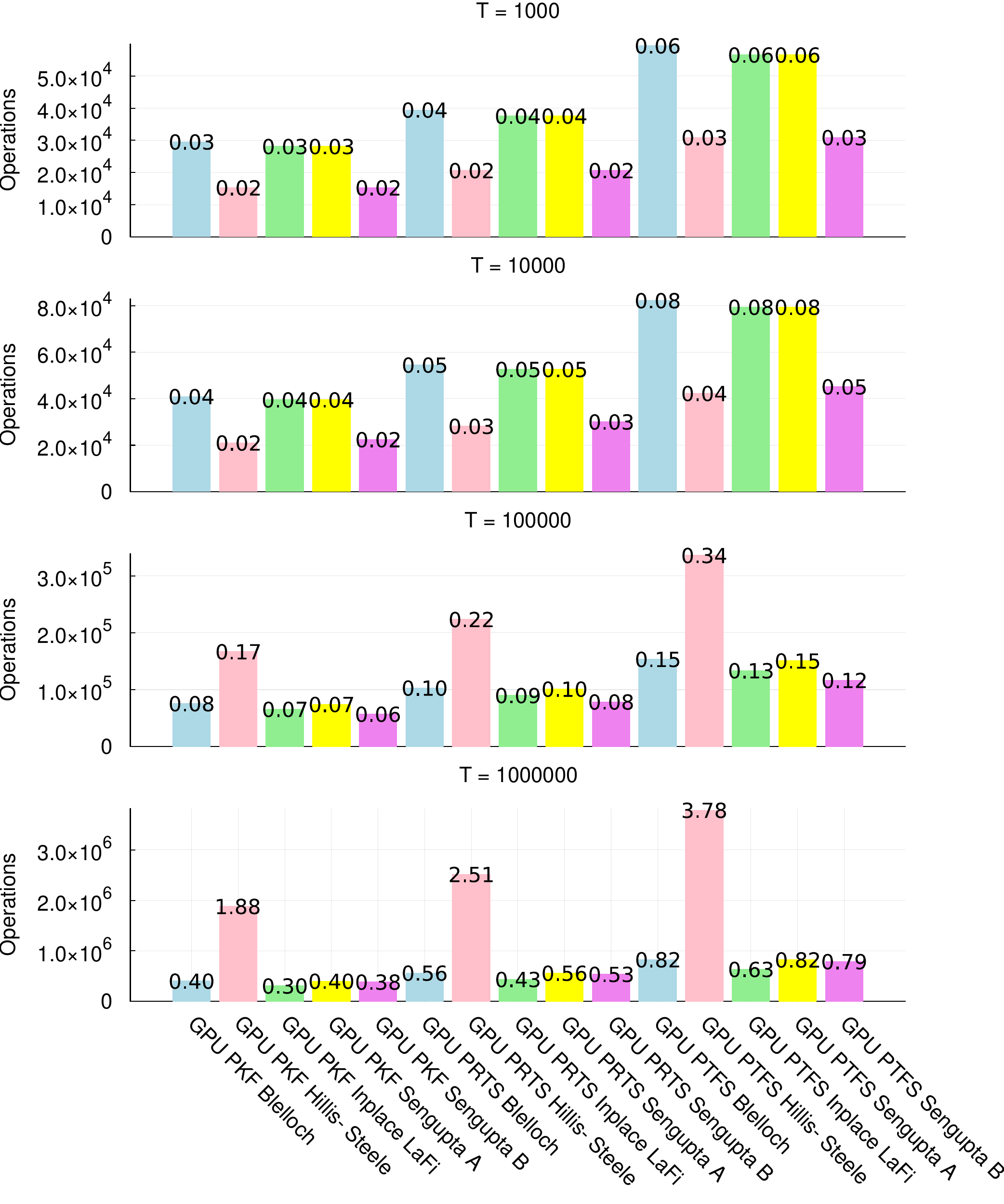}}
\caption{Number of operations of the different algorithm combinations with time series lengths $T = 10^3, 10^4, 10^5, 10^6$.}
\label{fig:pkfs_results_flops_snapshot}
\end{figure}

Figure~\ref{fig:pkfs_results_flops} shows the estimated time (in terms of number of operations) as a function of the number of time steps for the tested method, associative scan, and Kalman filter/smoother combinations. Figure~\ref{fig:pkfs_results_flops_snapshot} shows more detailed results of runs with $T = 10^3, 10^4, 10^5, 10^6$. It can be seen that with a small number of time steps $T$, the performance of the Hillis--Steele and Sengupta B (which incorporates Hillis--Steele) is the best. In particular, in Figure~\ref{fig:pkfs_results_flops_snapshot}, we can see that they are the fastest associative scan methods for $T = 10^3, 10^4$. However, at larger numbers of time steps ($T = 10^5, 10^6$ in Fig.~\ref{fig:pkfs_results_flops_snapshot}), the Hillis--Steele method is significantly slower than the other methods. At the larger numbers of time steps, the Inplace LaFi and Sengupta A appear to be slightly faster than the other associative scan methods.

Among the Kalman filtering and smoothing methods, the PKF appears to be the fastest systematically, followed by PRTS, and then PTFS. The good performance of PKF is, though, expected because PRTS and PTFS perform the same operations as PKF, along with additional ones.

\subsection{Speeds of different algorithms on Metal GPU} \label{sec:metal}

\begin{figure}[htbp]
\centerline{\includegraphics[width=\columnwidth]{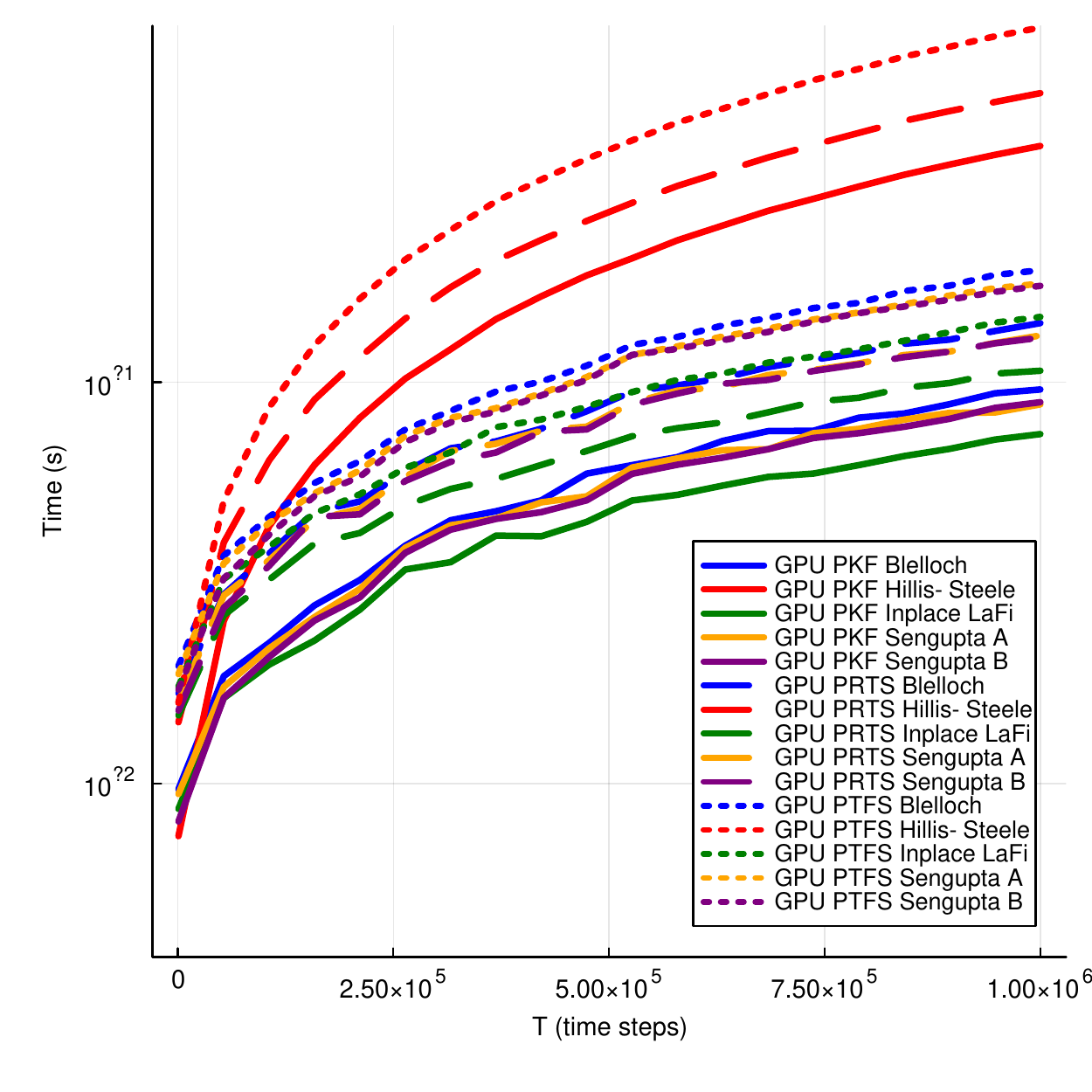}}
\caption{Run times of the different algorithm combinations on the Metal computer as a function of time series length $T$.}
\label{fig:pkfs_results_metal}
\end{figure}

\begin{figure}[htbp]
\centerline{\includegraphics[width=\columnwidth]{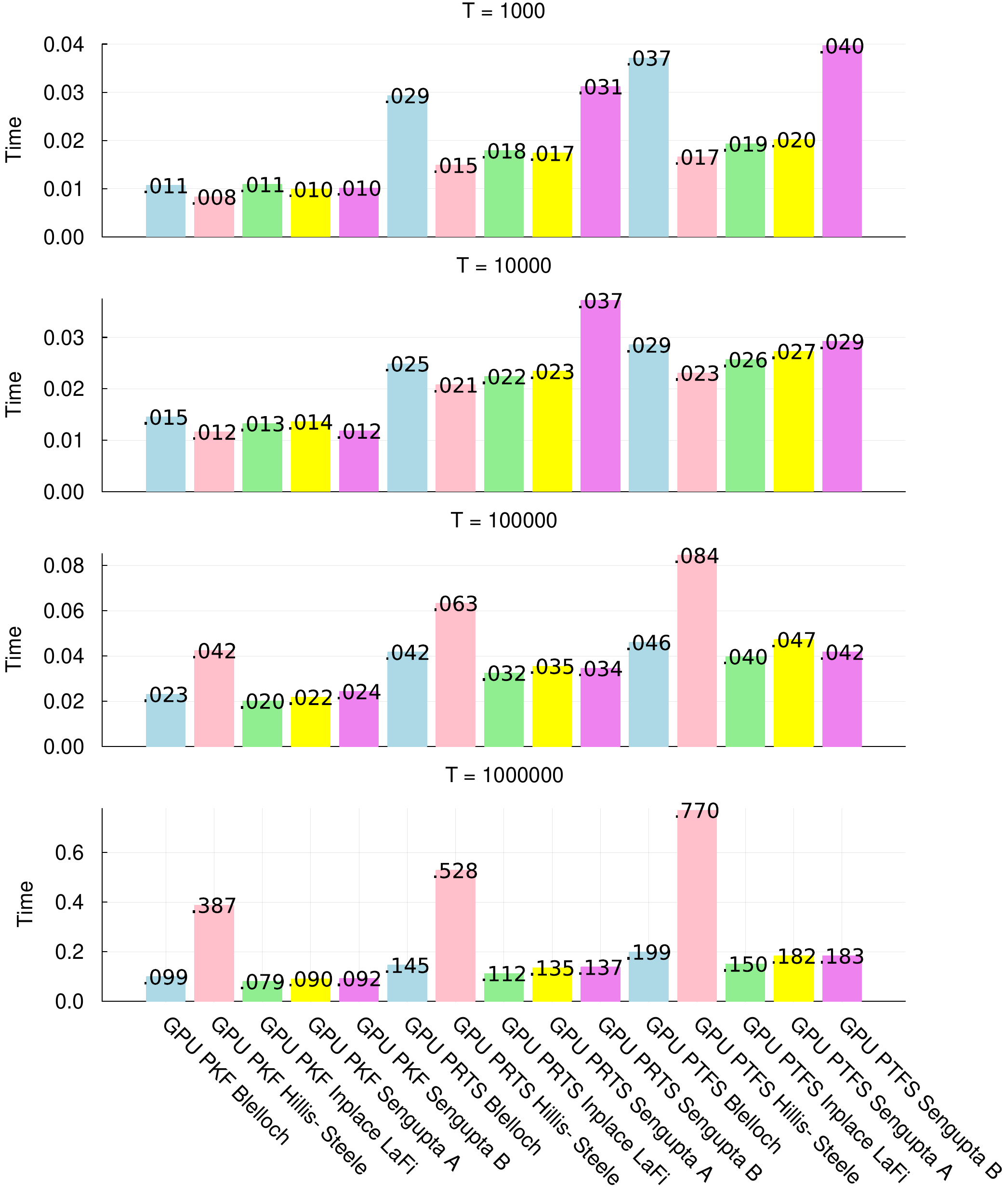}}
\caption{Run times of the different algorithm combinations on the Metal computer with time series lengths $T = 10^3, 10^4, 10^5, 10^6$.}
\label{fig:pkfs_results_metal_snapshot}
\end{figure}

We ran the same code that was used to implement the simulated hardware experiment in the previous section (Sec.~\ref {sec:simulated}), on the GPU of the Metal computer. Figure~\ref{fig:pkfs_results_metal} shows the run times as a function of the time series length $T$ and Figure~\ref{fig:pkfs_results_metal_snapshot} shows the times for time series lengths $T = 10^3, 10^4, 10^5, 10^6$. With short time series lengths ($10^3$, $10^4$), in the PKF case, all the associative scan methods perform similarly. However, in the PRTS and PTFS cases, Blelloch's method and Sengupta B seem to be slightly slower than the other methods. With the longer time series lengths ($10^5$, $10^6$), this effect disappears and all the methods except Hillis--Steele perform similarly, Hillis--Steele being significantly slower than the other methods. We can see that the Inplace LaFi method performs slightly better than the other methods with the longer time series lengths. We can see that the PKF method is always the fastest, PRTS the second fastest, and PTFS the third, as could be expected.

By comparing the simulated hardware results in Figure~\ref{fig:pkfs_results_flops_snapshot} and the Metal results in Figure~\ref{fig:pkfs_results_metal_snapshot}, we can observe that with the larger time series lengths, the simulation is a very good predictor for the actual time taken by the methods on the Metal GPU. However, with the shorter time series lengths, the simulation predicted Hillis--Steele and Sengupta B to be much better than they were on the actual hardware.

\subsection{Speeds of different algorithms on CUDA GPU} \label{sec:cuda}

\begin{figure}[htbp]
\centerline{\includegraphics[width=\columnwidth]{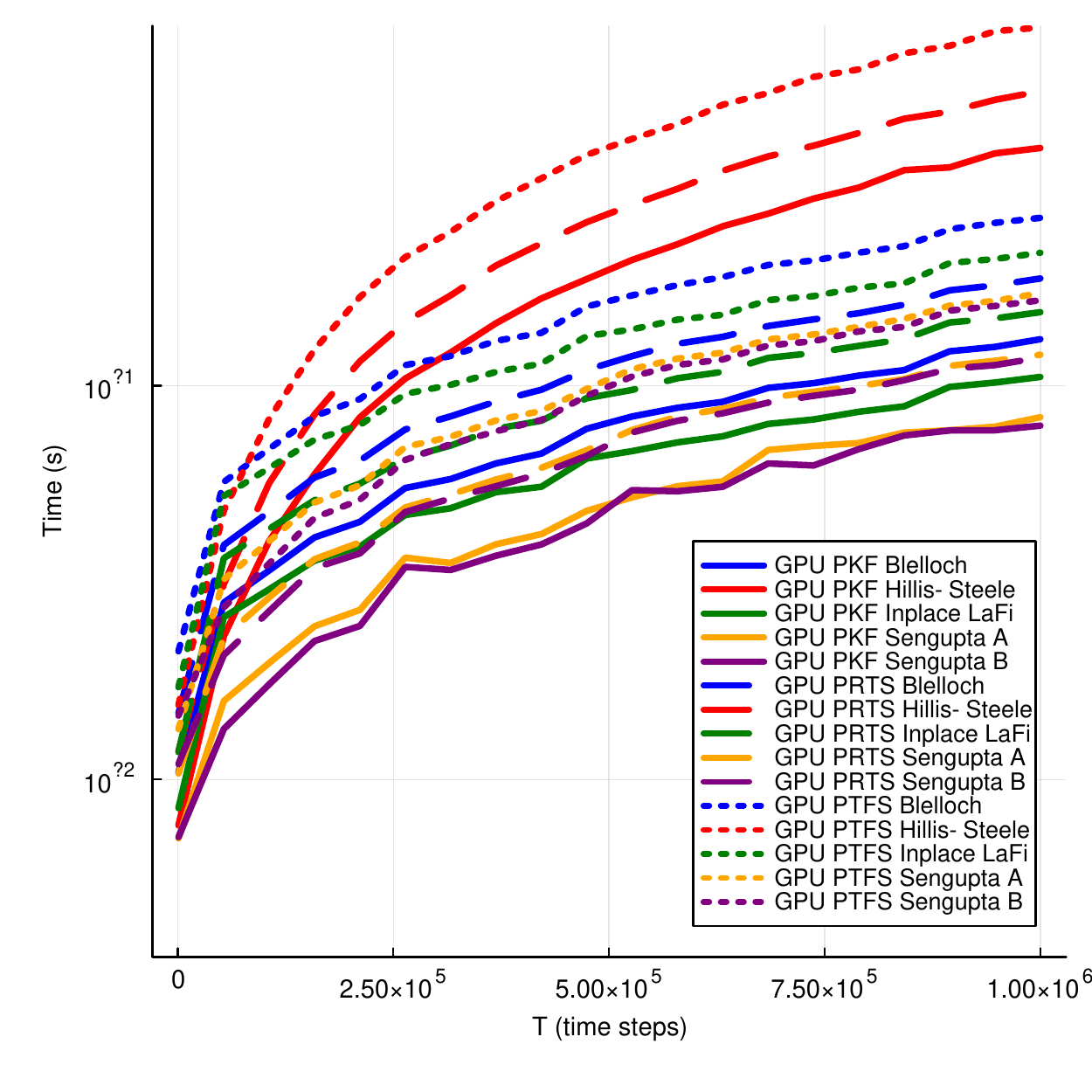}}
\caption{Run times of the different algorithm combinations on the CUDA computer as a function of time series length $T$.}
\label{fig:pkfs_results_cuda}
\end{figure}

\begin{figure}[htbp]
\centerline{\includegraphics[width=\columnwidth]{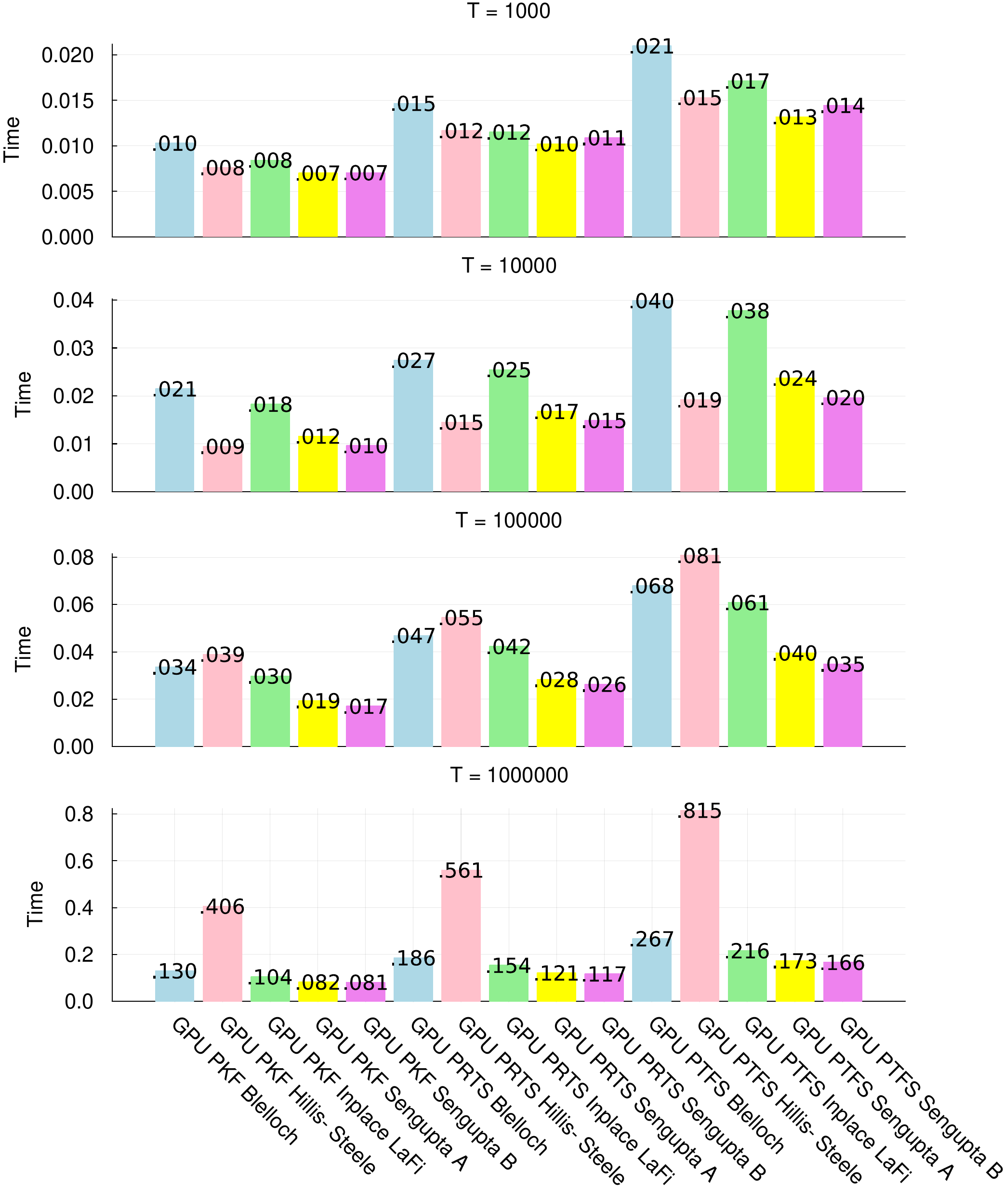}}
\caption{Run times of the different algorithm combinations on the CUDA computer with time series lengths $T = 10^3, 10^4, 10^5, 10^6$.}
\label{fig:pkfs_results_cuda_snapshot}
\end{figure}

We also ran the codes used in the simulated hardware experiment in Section~\ref{sec:simulated} and the Metal experiment in Section~\ref{sec:metal} on the GPU of the CUDA computer. The results are shown in Figures~\ref{fig:pkfs_results_cuda} and 
\ref{fig:pkfs_results_cuda_snapshot}. With the shorter time series lengths ($10^3$, $10^4$), we observe that Blelloch's method and Inplace LaFi are slightly slower than the other methods, whereas Hillis--Steele, Sengupta A, and Sengupta B perform well. With the longer time series lengths, the Hillis--Steele method starts to perform worse, and the differences between the other methods even out somewhat. However, Sengupta A and Sengupta B still perform the best. Again, the PKF methods are the fastest, then PRTS, and PTFS is the third fastest. 

When comparing to the simulated results in Figure~\ref{fig:pkfs_results_flops_snapshot}, we can see that the simulated results are again quite good predictors of the actual CUDA GPU behavior with the longer time series lengths, though not as good as in the case of the Metal GPU in Figure~\ref{fig:pkfs_results_metal_snapshot}. 

When comparing the Metal GPU results in Figure~\ref{fig:pkfs_results_metal_snapshot} to the CUDA results in Figure~\ref{fig:pkfs_results_cuda_snapshot}, we can see that the overall behavior of the GPUs is very similar. There are differences, though, which can be caused by the hardware differences or the differences in the code generated by the Julia GPU Compiler. However, it is surprising to see how similar the overall performance of the GPUs is. The codes that we run on both of the GPUs are exactly the same, and the computational times are both measured in the same way, which implies that the computational capabilities of the GPUs are very similar.

\subsection{Evaluation of PTFS on two GPUs}

In the previous sections, we have been running the algorithms on a single GPU. The results have also shown that in this setting, the PTFS algorithm which we proposed in Section~\ref{subsec:tf_smoother} is actually slower than the previously proposed PRTS \cite{Sarkka:2021}. However, as discussed in the same section, the forward and backward passes are independent and can be run on two GPUs in parallel. 

We implemented the PTFS algorithm described in Section~\ref{subsec:tf_smoother} on two CUDA GPUs as follows:
\begin{enumerate}
\item Run the backward pass of the PTFS on GPU0 and simultaneously the forward pass of the PTFS on GPU1.
\item Copy the results of the forward pass first to CPU, and from there to GPU0.
\item Then on GPU0 compute the combination step.
\end{enumerate}

\begin{figure}[htbp]
\centerline{\includegraphics[width=\columnwidth]{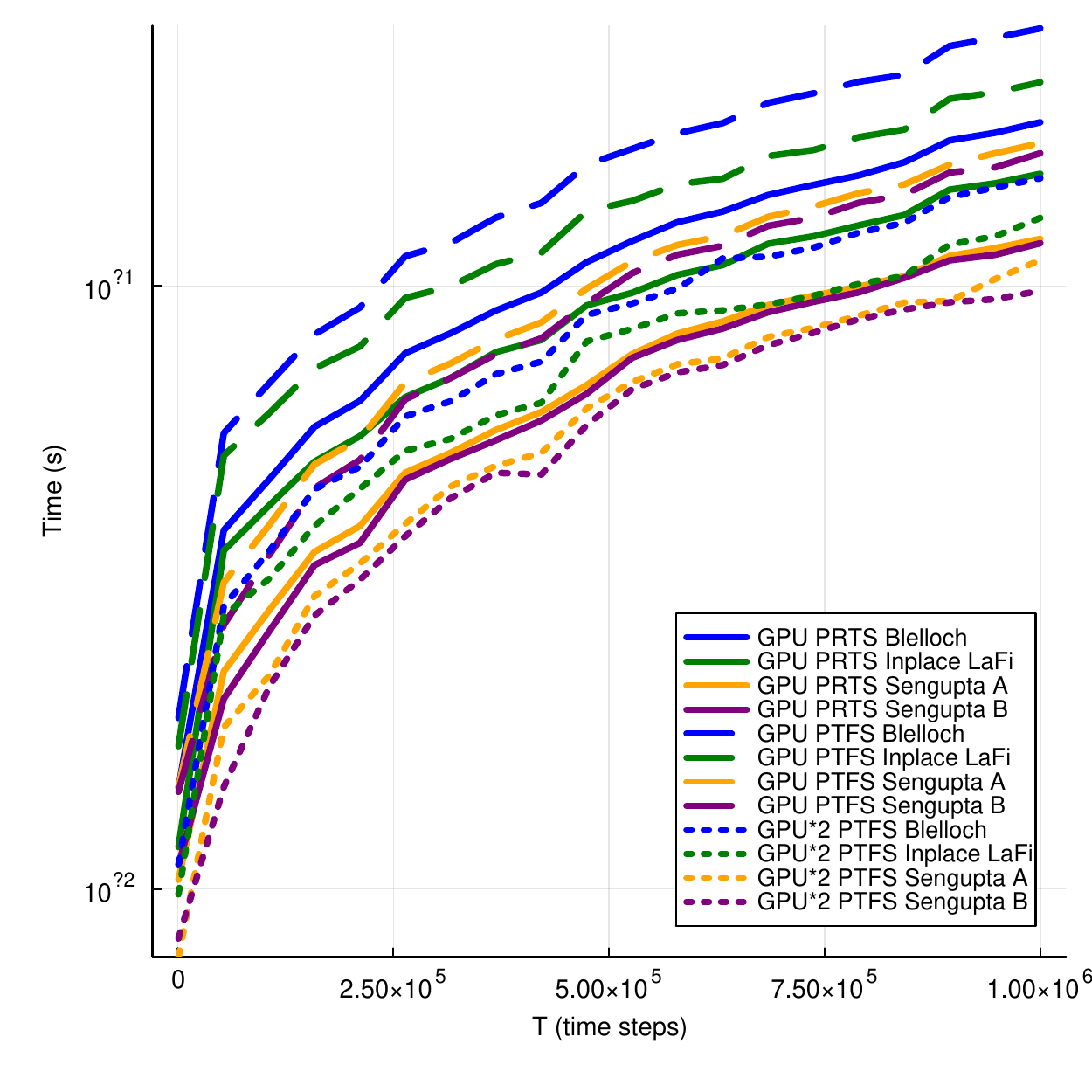}}
\caption{Run times of the 1-GPU versions of PRTS and PTFS, and the 2-GPU version of PTFS on the CUDA computer as a function of time series length $T$.}
\label{fig:tfs2_results_cuda}
\end{figure}

\begin{figure}[htbp]
\centerline{\includegraphics[width=\columnwidth]{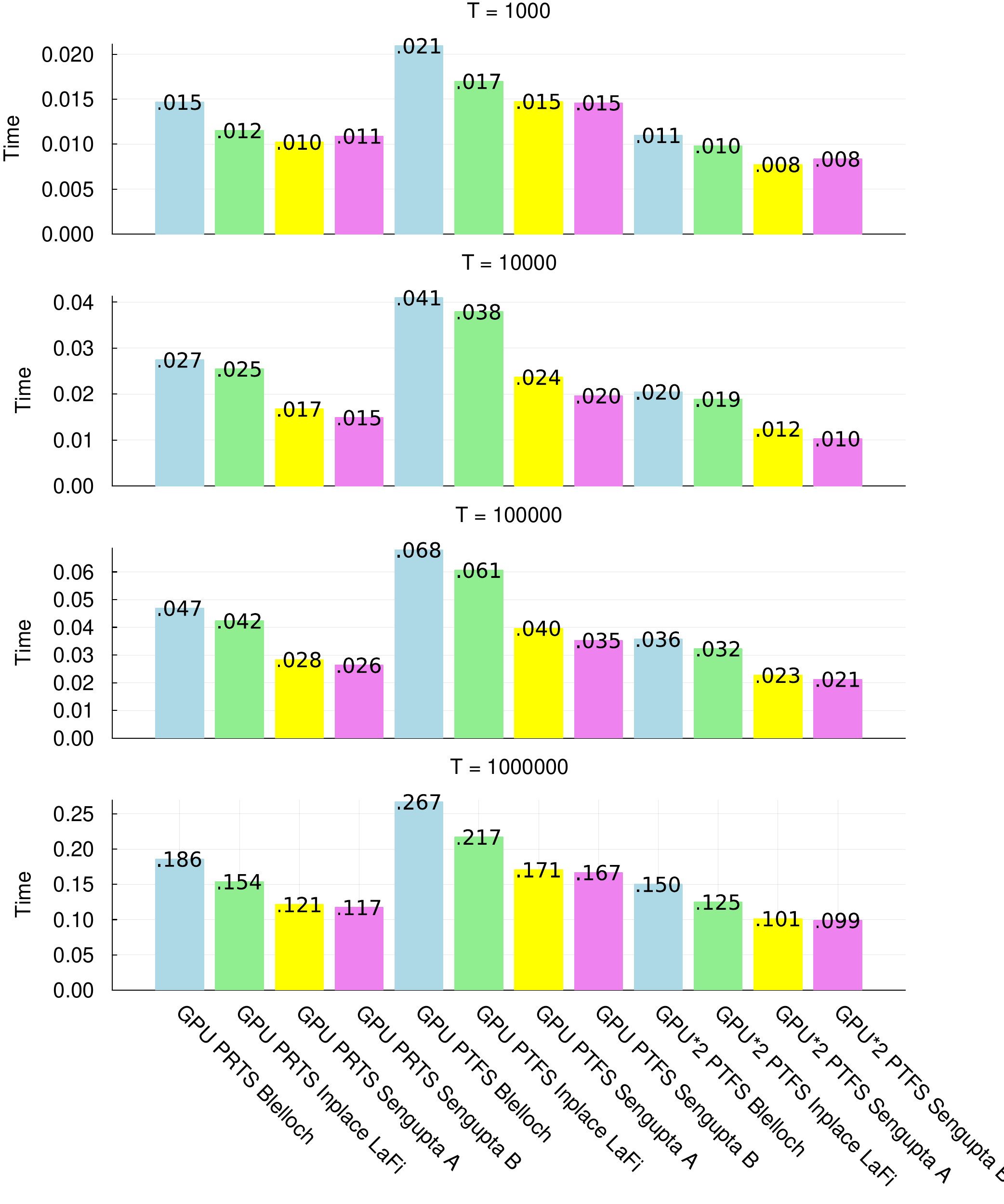}}
\caption{Run times of the 1-GPU versions of PRTS and PTFS, and the 2-GPU version of PTFS (denoted as GPU*2) on the CUDA computer with time series lengths $T = 10^3, 10^4, 10^5, 10^6$.}
\label{fig:tfs2_results_cuda_snapshot}
\end{figure}

The results of the experiment where we ran PRTS and PTFS methods on a single GPU and PTFS on two GPUs using the above procedure are shown in Figures~\ref{fig:tfs2_results_cuda} and \ref{fig:tfs2_results_cuda_snapshot}. The results now show that the 2-GPU version of the PTFS is significantly faster than either of the 1-GPU algorithms. Note that we have left out PKF because it is precisely the forward pass of PTFS, and we have also left out Hillis--Steele in order to see the differences of the methods more clearly. Among these tested methods, the 2-GPU PTFS with the Sengupta B associative scan appears to perform the best.

\subsection{GPU speedup analysis}

So far, we have only investigated the relative performance of the associative scan and Kalman filter/smoother combinations without computing the actual speedup that the methods provide. To compute this, we performed an experiment in which we ran the classical sequential Kalman filter \cite{Kalman:1960} (the optimal sequential algorithm) on a single core of a GPU and compared its speed to that of the parallel Kalman filter (PKF) \cite{Sarkka:2021}. We used the Inplace LaFi (Algorithm~\ref{alg:inplace-lafi}) as the associative scan algorithm.

\begin{figure}[htbp]
\centerline{\includegraphics[width=\columnwidth]{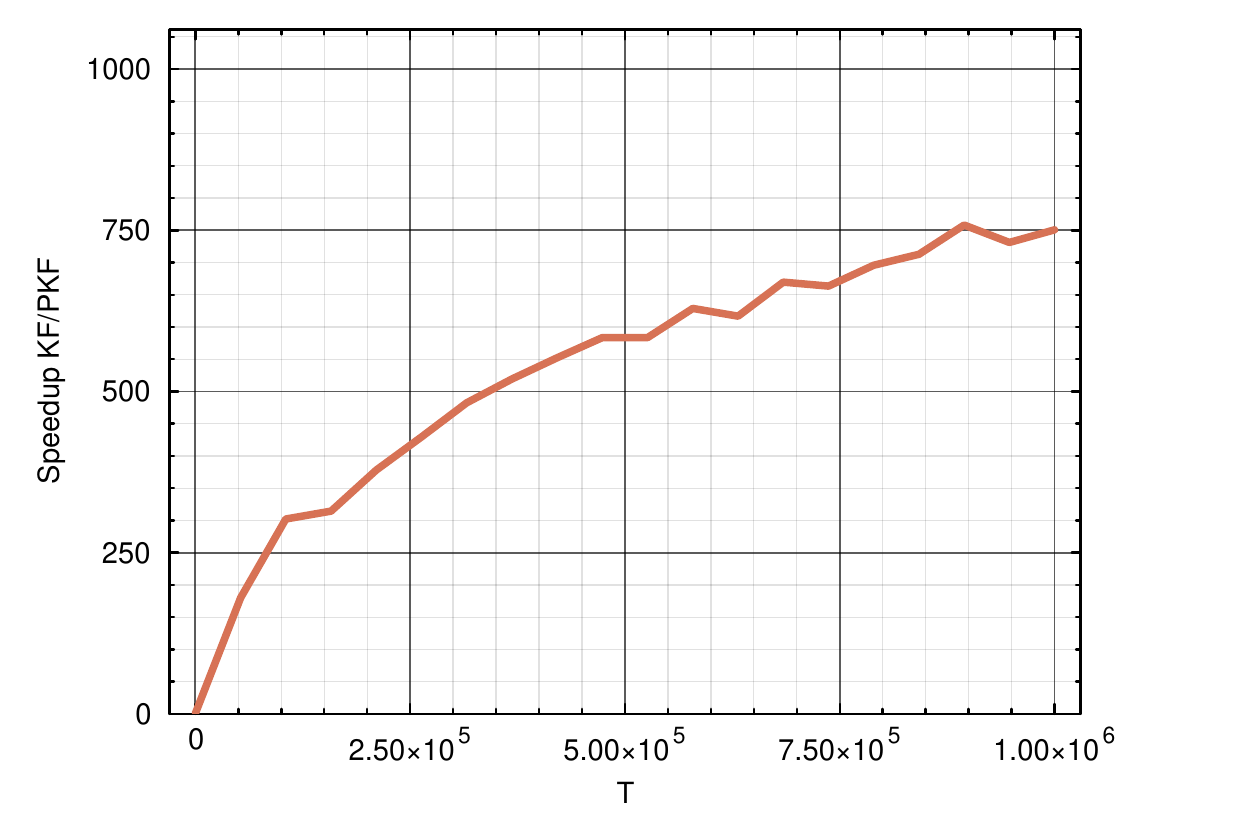}}
\caption{Speedup of the parallel Kalman filter on the Metal computer as a function of time series length $T$. The speedup increases from around 0 to a value close to 750.}
\label{fig:kf_results_metal_speedup}
\end{figure}

\begin{figure}[htbp]
\centerline{\includegraphics[width=\columnwidth]{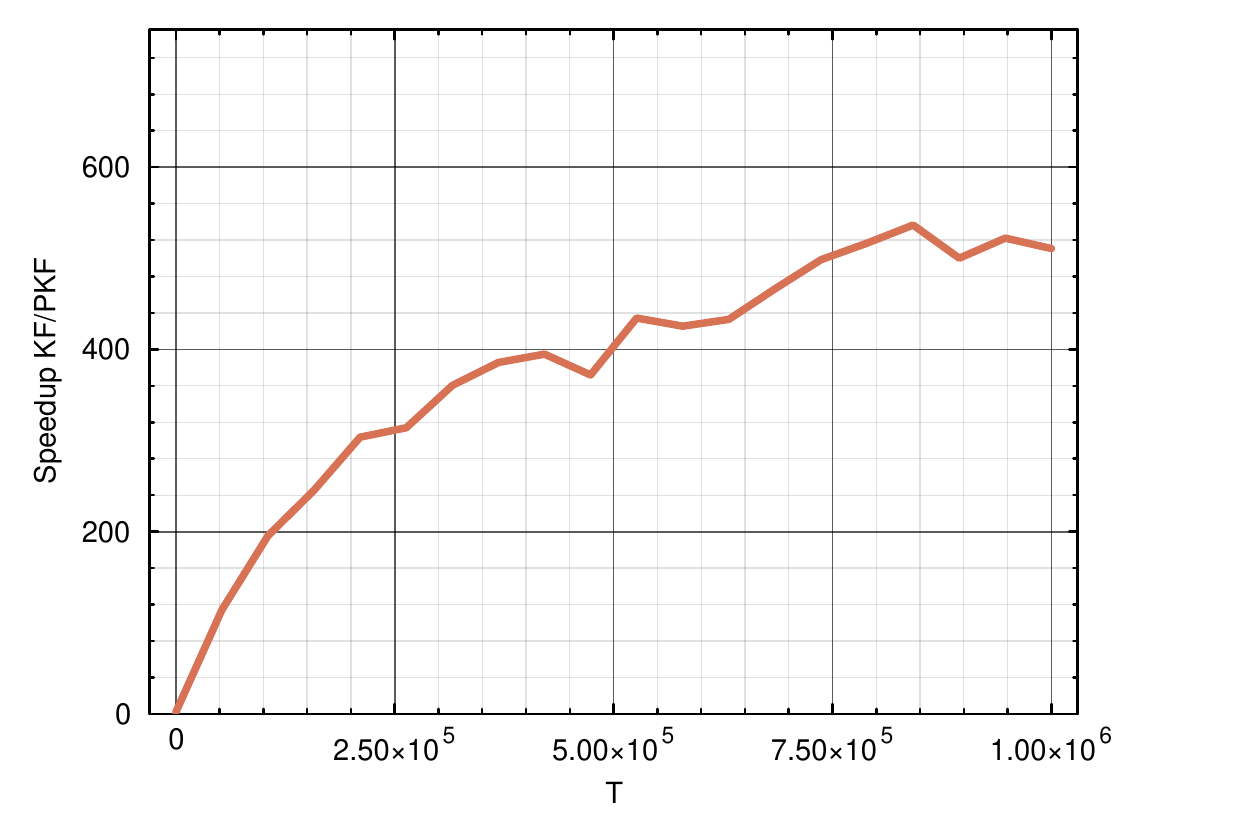}}
\caption{Speedup of the parallel Kalman filter on the CUDA computer as a function of time series length $T$. The speedup increases from around 0 to a value close to 500.}
\label{fig:kf_results_cuda_speedup}
\end{figure}

The speedups (= ratios of sequential and parallel times) for Metal GPU and CUDA computers are shown in Figures~\ref{fig:kf_results_metal_speedup} and \ref{fig:kf_results_cuda_speedup}, respectively. It can be seen that the speedup on the Metal GPU increases up to around 750, and the speedup in the CUDA GPU to around 500. 

\section{Conclusion} \label{sec:conclusion}
In this paper, we have performed an experimental evaluation of temporally parallel Kalman filters and smoothers using both simulated and real GPU hardware. A novel parallel two-filter smoother was also proposed, and the Julia codes for the Metal and CUDA GPU experiments are made publicly available.

The experimental results show that the work complexity of the parallel scan method has a significant effect on he practical performance of the methods. In particular, the Hillis--Steele-based methods exhibit the worst performance among the tested parallel scan methods when the time series length is significant, whereas the other methods perform more evenly. The simulated GPU hardware appears to be a very good predictor of the performance of the methods on real GPU hardware. Furthermore, the performances of the methods on Metal and CUDA GPUs are very similar.

The experimental results also show that while the proposed parallel two-filter smoother (PTFS) is slower than the parallel Rauch-Tung-Striebel (PRTS) smoother on a single GPU, it outperforms the PRTS on a two-GPU setup.

\section*{Acknowledgments}
The authors would like to thank the Research Council of Finland for funding.

\bibliographystyle{IEEEtran}
\bibliography{IEEEabrv,gpu-kalman-paper-2025}

\begin{IEEEbiography}[{\includegraphics[width=1in,height=1.25in,clip,keepaspectratio]{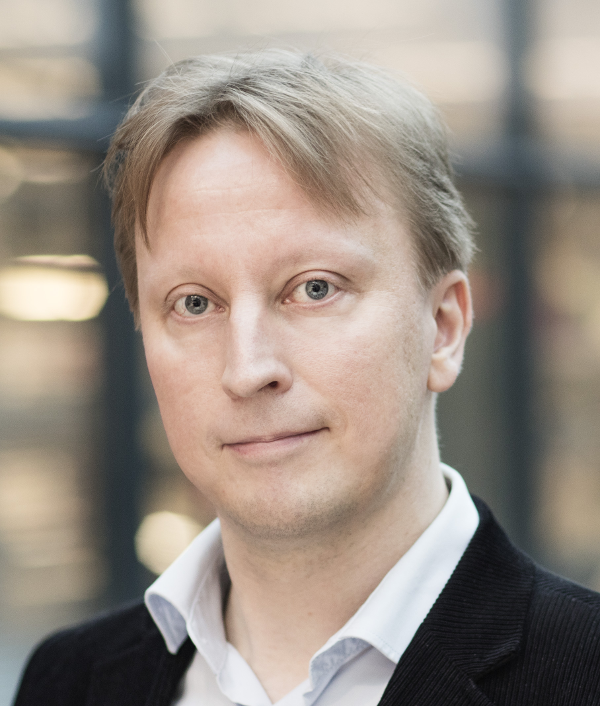}}]{Simo S\"arkk\"a} received his MSc.\ degree (with distinction) in engineering physics and mathematics, and DSc.\ degree (with distinction) in electrical and communications engineering from Helsinki University of Technology, Espoo, Finland, in 2000 and 2006, respectively. Currently, he is a Professor with Aalto University. His research interests are in multi-sensor data processing and control systems. He has authored or coauthored over 200 peer-reviewed scientific articles and three books. He is a Senior Member of IEEE.  \end{IEEEbiography}

\begin{IEEEbiography}[{\includegraphics[width=1in,height=1.25in,clip,keepaspectratio]{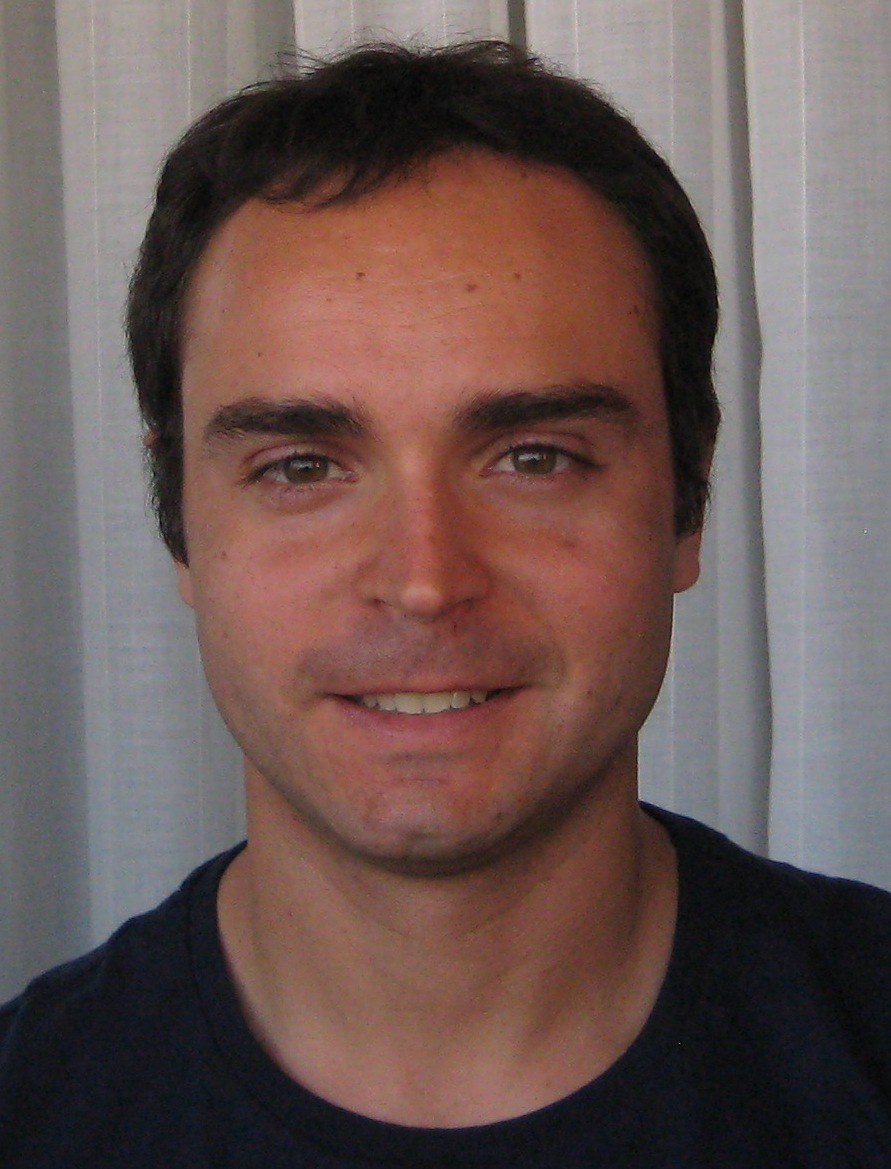}}]{\'Angel F. Garc\'ia-Fern\'andez} received the telecommunication engineering degree and the Ph.D. degree from Universidad Polit\'ecnica de Madrid (UPM), Spain, in 2007 and 2011, respectively. \\ 
He is currently an Associate Professor at the Information Processing and Telecommunications Center of UPM. Previously, he held other academic positions at Chalmers University of Technology, Sweden, Curtin University, Australia, Aalto University, Finland, and the University of Liverpool, UK. His main research interests are in the area of Bayesian inference, with emphasis on multiple target tracking.
\end{IEEEbiography}

\vfill

\end{document}